\documentclass{IEEEtran}

\usepackage{algorithmic}
\usepackage{textcomp}

\usepackage{chngcntr}
\usepackage{url}
\usepackage{amsmath,stackengine}
\stackMath
\usepackage{amsfonts}%
\usepackage{amssymb}%
\usepackage{bm}%
\usepackage{graphicx}
\usepackage{caption}
\usepackage{subfigure}
\usepackage{cite}
\usepackage{multirow}
\usepackage{boldline}
\usepackage{enumitem}
\usepackage{nomencl}
\makenomenclature
\usepackage{cleveref}
\usepackage{tikz}
\usetikzlibrary{automata,positioning}
\usetikzlibrary{arrows,automata}
\tikzset{
  font={\fontsize{7pt}{12}\selectfont}}
\usepackage{lipsum}
\usepackage{float}
\usepackage{mathtools}

\usepackage[english]{babel}
\usepackage[autostyle]{csquotes}
\usepackage{array,booktabs}
\newcolumntype{P}[1]{>{\centering\arraybackslash}p{#1}}
\newcolumntype{M}[1]{>{\centering\arraybackslash}m{#1}}

\makeatletter
\def\@seccntformat#1{\@ifundefined{#1@cntformat}%
   {\csname the#1\endcsname\quad}
   {\csname #1@cntformat\endcsname}
}
\makeatother

\newtheorem{theorem}{Theorem}

\newtheorem{assumption}{Assumption}

\newtheorem{corollary}{Corollary}

\newtheorem{definition}{Definition}

\newtheorem{lemma}{Lemma}

\newtheorem{proposition}{Proposition}
\newtheorem{remark}{Remark}

\def\BibTeX{{\rm B\kern-.05em{\sc i\kern-.025em b}\kern-.08em
    T\kern-.1667em\lower.7ex\hbox{E}\kern-.125emX}}
\begin{document}
\title{Decentralized LQ-Consistent Event-triggered Control over a Shared Contention-based Network}
\author{M. Balaghiinaloo\textsuperscript{1},~\IEEEmembership{Student member,~IEEE}, D. J. Antunes\textsuperscript{1},~\IEEEmembership{Member,~IEEE}, \\M. H. Mamduhi\textsuperscript{2},~\IEEEmembership{Member,~IEEE}, S.~Hirche\textsuperscript{3},~\IEEEmembership{Fellow,~IEEE}
\thanks{\noindent\textsuperscript{1}Hadi Balaghiinaloo and Duarte J. Antunes are with the Control Systems Technology Group, Department of Mechanical Engineering, Eindhoven University of Technology, P.O.~Box 513, NL-5600 MB, Eindhoven,
        The Netherlands e-mail: {\tt\small \{m.balaghiinaloo, d.antunes\}@tue.nl}. }
\thanks{\noindent\textsuperscript{2} Mohammad H. Mamduhi is with the Division of Decision and Control Systems, KTH Royal Institute of Technology, SE-100 44, Stockholm, Sweden  e-mail: {\tt\small mamduhi@kth.se}. }
\thanks{\noindent\textsuperscript{3} Sandra Hirche is with the Chair for
Information-oriented Control, Technical University of Munich, Arcisstra\ss e 21, D-80290 Munich, Germany e-mail: {\tt\small hirche@tum.de}. }

\thanks{\noindent This project was funded from the European Union's Horizon 2020 Framework Program for Research and Innovation
under grant agreement No~674875 and DFG Priority Program SPP1914 "Cyber-physical Networking".}}

\maketitle

\begin{abstract}
Consider a network of multiple independent stochastic linear systems where, for each system, a scheduler collocated with the sensors arbitrates data transmissions to a corresponding remote controller through a shared contention-based communication network. While the systems are physically independent, their optimal controller design problems may, in general, become coupled, due to network contention, if the schedulers trigger transmissions based on state-dependent events. In this article we propose a class of probabilistic admissible schedulers for which the optimal controllers, with respect to local standard LQG costs, have the certainty equivalence property and can still be determined decentrally. Then, two scheduling policies within this class are introduced; a non-event-based and an event-based, both with an easily adjustable triggering probability at every time step. We then prove that, for each closed-loop system, the event-based scheduler and its optimal controller outperforms the closed-loop system with the non-event-based scheduler and its associated optimal controller. Moreover, we show that, for each closed-loop system, the optimal state estimators for both scheduling policies follows a linear iteration. Finally, we provide a method to regulate the triggering probabilities of the schedulers by maximizing a network utility function.
\end{abstract}


\section{Introduction}
\par
\begin{figure}[!t]
    \centering
    \includegraphics[width=8cm]{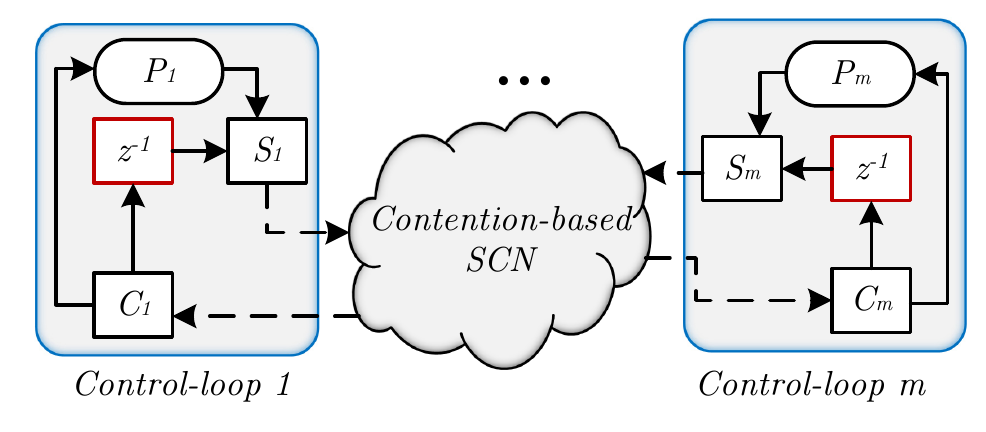}
    \caption{Networked control loops through a shared communication network (SCN); \textit{P} describes the plant, \textit{S} the scheduler, \textit{C} the controller and $z^{-1}$ a one time step delay.}
    \label{fig:plant}
\end{figure}

 Event-triggered control (ETC) pertains to strategies that manage transmissions in a control loop based on events rather than time and it is intended for scenarios where communication resources are scarce or costly. Extensive research has been carried out in the past decade on ETC, mostly focusing on  a single control loop closed over a single communication channel~\cite{heemels2012introduction, antunes2014rollout,soleymani2017event, nowzari2017event, Asadi2018Aconsistent}. However, communication management is especially interesting in settings where the communication network is shared by multiple control loops~\cite{wang2011event, molin2014bi, mamduhi2014event}, as illustrated in Fig.~\ref{fig:plant}.  Contention-free protocols such as time-division multiple access (TDMA)~\cite{kurose2005computer} enable periodic transmission for all the control loops but lead to inefficient bandwidth usage when some of the control loops alternate between being active and inactive. In fact, under these conditions, systems operating through these protocols need a central coordinator for the resource reallocation between different users, hence decentralized scalability is an issue for these medium access control protocols. In turn, contention-based protocols such as slotted-ALOHA~\cite{kurose2005computer} are decentrally reconfigurable. However, contention in these protocols could hamper the control design and analysis, when the loops transmit based on events. In fact, if event-based data triggering is influenced by the states of the systems and the control inputs, then due to network contention, every local optimal controller depends on the state of all network dynamic users and this hampers  decentralization.
\par \textit{Contributions:} In this article, we first propose a class of decentralized admissible schedulers for each control loop in a contention-based setting, within which the design of optimal control strategies can be performed in a decoupled fashion. This class specifies that transmissions occur based on a function of primitive random variables, i.e., disturbances, noise and any other possible independent random variable. Then we consider a subclass of these policies for which the triggering probability can easily be tuned. Naturally, a non-event-based purely stochastic transmission~(PST) policy belongs to this class. Our main contribution is to propose an event-triggered controller, consisting of an event-based scheduling policy in this class and an optimal control policy with respect to an average quadratic cost, that outperforms the PST policy also with the associated optimal control input. Inspired by~\cite{Brunner2018stochastic}, we refer to this property as \textit{LQ-consistency}. Moreover, we show that, for each system, the local optimal state estimator for both scheduling policies follows a linear iteration. Finally, we propose a method to regulate the schedulers based on their triggering probabilities in order to maximize a network utility function in the sense of providing proportional fairness between the users and taking into account the control performance of each loop.
\par \textit{Related work}: The event-based scheduler proposed in this work is inspired by the stochastic threshold event-triggered transmission (STETT) policy \cite{han2017stochastic,Brunner2018stochastic}, based on which data is triggered when the norm of the error between local and remote state estimators becomes larger than an exponentially distributed random threshold.
A similar decentralized networked control structure, as shown in Fig.~\ref{fig:plant}, is also investigated in~\cite{gatsis2016control}, and sufficient conditions are given for the scheduling policies in order to guarantee the mean square stability (MSS) of all the network users. Furthermore, \cite{gatsis2016control} proposes an innovation-based deterministic event-triggered scheduler, for which the triggering probability is regulated by assuming that the state vector always follows a Gaussian distribution with a known covariance. However, this is a simplifying assumption and typically, deterministic event-triggered schedulers destroy the Gaussianity of the state~\cite{han2017stochastic}.
\par Comparing the performance of time-triggered and event-triggered scheduling policies over shared communication networks has received considerable attention in recent years~\cite{cervin2008scheduling,blind2013time,xia2017networked}. In~\cite{cervin2008scheduling}, it is shown, via a numerical example, that a threshold-based event-triggered scheduling policy with a CSMA protocol results in a better performance, measured by the average state variance, than a time-triggered policy with TDMA protocol. Moreover, \cite{blind2013time} considers an average state variance performance index and compares its values for different communication protocols and load conditions.
In addition, \cite{xia2017networked} claims that event-triggered policies for state estimation may perform worse than time-triggered policies, in the sense of average state error covariance, if the contention-resolution mechanism of the communication network is explicitly considered. Some researchers also consider stability of the control loops when transmitting through a shared communication network~\cite{Ramesh2011ACC,mamduhi2016decentralized,mamduhi2017error}.  For instance, in~\cite{Ramesh2011ACC}, the triggering probabilities of every scheduler transmitting through a network with a CSMA protocol are determined in order to guarantee Lyapunov MSS. These triggering probabilities can be used to tune the threshold of event-based scheduling policies. 
An optimal co-design problem of an event-triggered scheduler and a controller is investigated in~\cite{molin:13},  establishing the optimality of the linear certainty equivalent controller when the event-based scheduler only depends on the primitive random variables, see also~\cite{goldenshluger2017minimum}. Note that we extend the result of~\cite{molin:13} for multiple control loops closed over a shared contention-based network. 
\par \textit{Organization}: The remainder of this article is organized as follows: the problem of interest and the class of admissible schedulers are introduced in Section~\ref{sec:problemsetting}. The decentralized optimal control policy for the admissible schedulers is determined in Section~\ref{sec:decentralizedcontrol}. In Section~\ref{sec:Purelystochastic}, the non-event-based PST policy is discussed. The novel event-based scheduling policy is introduced in Section~\ref{sec:proposedpolicy} and the main results are presented in Section~\ref{sec:mainresults}. Moreover, network utility maximization is discussed in Section~\ref{sec:regulation}. The presented results are validated through a numerical example in Section~\ref{sec:Numericalresult} and Section~\ref{sec:conclusion} presents some concluding remarks. 
\par \textit{Notation}: \mbox{$\mathsf{f}(x|y)$} denotes the conditional probability density function (pdf) of a random variable $x$ given the information set $y$ and \mbox{$\mathsf{N}(\bar{y},Y)$} indicates a multi-variate Gaussian pdf with mean $\bar{y}$ and covariance $Y$. The probability of event $x$ is denoted by $\mathsf{Pr}(x)$; \mbox{$\delta\sim\mathsf{B}(p)$} indicates that the random variable $\delta$ follows a Bernoulli distribution with success probability~$p$. By \mbox{$\varrho(A)$} and \mbox{$\text{tr}(A)$} we denote the spectral radius and the trace of the square matrix~$A$, respectively. Moreover, \mbox{$\mathbb{N}_0:=\mathbb{N}\cup \{0\}$} in which $\mathbb{N}$ is the set of natural numbers, \mbox{$Z_t^s:=\{k\in\mathbb{Z}|t\leqslant k\leqslant s\}$}, where $\mathbb{Z}$ is the set of integers, and $a_{0:k}=\{a_i\in\mathbb{R}^n|i\in Z_0^k,~n\in\mathbb{N}\}$.
\section{Problem setting}\label{sec:problemsetting}
Consider a networked control system (NCS) comprised of multiple independent stochastic linear time-invariant (LTI) subsystems each modeled by the following dynamics
\begin{equation}\label{eq:plant}
\begin{aligned}
x^i_{k+1}& = A_ix^i_k+B_i u^i_k + w^i_k,~~~~
y^i_{k} =C_ix^i_k+v^i_k,
\end{aligned}
\end{equation}
in which \mbox{$x^i_k\in\mathbb{R}^{n_i}$}, \mbox{$u^i_k\in \mathbb{R}^{m_i}$} and \mbox{$y^i_k\in\mathbb{R}^{o_i}$} are, respectively, the state, the control input and the output vectors at time step \mbox{$k \in \mathbb{N}_0$}, for every \mbox{$i\in Z_1^m$}, where $m$ is the total number of subsystems. Let \mbox{$\{w^i_k|k\in\mathbb{N}_0\}$} and \mbox{$\{v^i_k|k\in\mathbb{N}_0\}$} be sequences of independent and identically distributed (i.i.d.) Gaussian random variables with zero means and positive definite covariances \mbox{$W_i = \mathsf{E}[w^i_kw^{i\intercal}_k]$}, and \mbox{$V_i = \mathsf{E}[v^i_kv^{i\intercal}_k]$}, for every \mbox{$k\in \mathbb{N}_0$}. Moreover, the pairs \mbox{$(A_i,B_i)$} and \mbox{$(A_i,C_i)$} are assumed to be controllable and observable, respectively. The performance of each subsystem is measured by the following local average quadratic cost
\begin{equation}\label{eq:per}
J^i:= \limsup_{T\rightarrow \infty}\frac{1}{T}\mathsf{E}[\sum_{k=0}^{T-1}x_k^{i\intercal} Q_ix^i_k + u_k^{i\intercal} R_iu^i_k],
\end{equation}
in which $Q_i$ and $R_i$ are positive semi-definite and positive definite matrices with appropriate dimensions, respectively, and \mbox{$(A_i,Q^{\frac{1}{2}}_i)$} is assumed to be observable for every \mbox{$i\in Z^m_1$}. Therefore, each subsystem is characterized by the tuple $(A_i,B_i,C_i,W_i,V_i,Q_i,R_i)$, which in general is different from the other subsystems, i.e., they are heterogeneous.
\par As depicted in Fig. \ref{fig:plant}, we assume that the sensors of every subsystem are collocated with a scheduler that arbitrates data transmissions in the control loop over the shared communication network. Moreover, since different subsystems are assumed to be physically independent, every scheduler just needs to transmit its locally  acquired information to its corresponding remote controller. We also assume time-synchronization of the sampling process of different subsystems, where they all have equal sampling periods. Besides the performance index~\eqref{eq:per} for each individual control loop, one may also define a social quadratic cost such as \mbox{$J=\sum_{i=1}^{m}J^i$} or a network utility function as suggested in~\cite{lee2006jointly}. This will be discussed later in Sections~\ref{sec:Purelystochastic} and~\ref{sec:regulation}.
\subsection{Assumptions on the shared communication network}\label{sec:MAC}
\par The assumptions on the shared communication network
are very close to the ones considered in the context of the contention-based protocols \cite{kurose2005computer} and
can be summarized as follows (more explanations in~\cite{balaghi2018decentralized}):
\begin{itemize}
  \item Time is partitioned into fixed-size slots; during each time-slot only one user can transmit successfully.
  \item All subsystems are restricted to start a new transmission at the beginning of the time-slot.
  \item Collision will occur if more than one user attempt to transmit, which results in loss of the collided data.
  \item Every scheduler receives a data receipt acknowledgment at the same time-slot if data transmission is successful, otherwise, it is assumed that a collision has occurred.
  \item There is no retransmission of the collided data.
  \item The transmission time is assumed to be negligible with respect to the duration of the time-slot.
\end{itemize}
\subsection{Information structure}
In this section, we introduce the information set available for the scheduler and the controller at every time step. Let
$$\delta^i_k=\begin{cases}
               1, & \mbox{if the scheduler $i$ attempts data transmission at $k$}  \\
               0, & \mbox{otherwise},
             \end{cases}$$
and
$$\rho^i_k=\begin{cases}
               1, & \mbox{if the network is available for the user \mbox{$i$} at $k$}  \\
               0, & \mbox{otherwise},
             \end{cases}$$
for every user \mbox{$i\in Z^m_1$}, at every time step \mbox{$k\in\mathbb{N}_0$}. Based on the properties of the shared contention-based communication network introduced in Section \ref{sec:MAC}, we can write
$\rho^i_k=\prod_{j=1,j\neq i}^{m}(1-\delta^j_k).$
Moreover, let
$\sigma^i_k = \rho^i_k\delta^i_k$ be a variable indicating a successful transmission at every time step, in which case \mbox{$\sigma^i_k=1$}, and \mbox{$\sigma^i_k=0$}, otherwise.
Note that based on the structure of the shared communication network in Section \ref{sec:MAC}, every scheduler receives an error-free acknowledgment signal from the controller whenever an attempted transmission is successful, and therefore, it knows all the previous values of \mbox{$\rho^i_{k}$}. Accordingly, in order to decide on \mbox{$\delta^i_k$} the scheduler has the following information set at every time step $k$,
\begin{equation}\label{eq:info-set-s}
\mathsf{L}^i_k:=\{\delta^i_\ell,y^i_\ell|\ell \in Z_0^{k-1}\}\cup\{\rho^i_\ell|\delta^i_\ell=1\land \ell \in Z_0^{k-1}\}\cup \{y^i_k\}.
\end{equation}
\par Let \mbox{$\hat{x}^i_{k|k}=\mathsf{E}[x^i_k|\mathsf{L}^i_k]$} and \mbox{$\hat{x}^i_{k+1|k}=\mathsf{E}[x^i_{k+1}|\mathsf{L}^i_k]$} be the state estimations of the current and the next time step computed by the local state estimator collocated with the scheduler.
We know that the optimal least-square state estimator at the scheduler follows a linear iteration and can be computed recursively by the Kalman filter as follows
\begin{equation}\label{eq:scheduler-estimate}
 \begin{aligned}
 \hat{x}^i_{k+1|k}&=A_i\hat{x}^i_{k|k}+B_iu^i_k,&\\
 \hat{x}^i_{k|k}&=\hat{x}^i_{k|k-1}+L_i(y^i_k-C_i\hat{x}^i_{k|k-1}),
 \end{aligned}
 \end{equation}
 where~$ L_i=\bar{\Theta}_i C_i^{\intercal}(C_i\bar{\Theta}_i C_i^{\intercal}+V_i)^{-1}$, and
 $$\begin{aligned}
 \bar{\Theta}_i&=A_i\Theta_i A_i^\intercal +W,~~~~
 \Theta_i =\bar{\Theta}_i-L_i(C_i\bar{\Theta}_i C_i^{\intercal}+V_i)L_i^\intercal.
 \end{aligned}
$$ 
 For simplicity we assume that $x_0^i$ is a Gaussian random variable with zero mean and covariance $\Theta_i$ ($\mathsf{E}[(x^i_0-\hat{x}^i_{0|0})(x^i_0-\hat{x}^i_{0|0})^\intercal|\mathsf{L}^i_0]=\Theta_i$), which implies that $\mathsf{E}[(x^i_k-\hat{x}^i_{k|k})(x^i_k-\hat{x}^i_{k|k})^\intercal|\mathsf{L}^i_k]=\Theta_i,~\forall k\in \mathbb{N}_0$ since \mbox{$\Theta_i$} is the fixed point of the Kalman filter's time-varying Ricatti equation.
\par When the triggering condition is satisfied, i.e., \mbox{$\delta_k^i=1$}, 
the scheduler transmits \mbox{$\hat{x}^i_{k|k}$} to the controller. 
Accordingly, the information set available for the controller at every time step \mbox{$k\in\mathbb{N}_0$} is as follows
\begin{equation}\label{eq:info-set-c}
\mathsf{H}^i_k:=\{\hat{x}^i_{\ell|\ell}| \sigma^i_\ell=1\land \ell \in Z_0^{k}\}\cup\{\sigma^i_\ell | \ell \in Z_0^{k}\}.
\end{equation}
\subsection{Required characteristics for the scheduling policies}
\par We define next a class of admissible schedulers for which not only the optimal controller has the certainty equivalence property, but also it can be computed decentrally for the dynamic users of the shared contention-based communication network, as it will be discussed in Section~\ref{sec:decentralizedcontrol}. Let us denote all independent random variables of any control loop, such as $x^i_0$, $w^i_k$, $v^i_k$ at all time steps, as the primitive random variables.
\begin{definition}\label{def:admissiblepolicy}
(\textit{Admissible schedulers})
Suppose \mbox{$a^i_{\ell}\in\mathbb{R}^{r_i}$} for all \mbox{$\ell\in\mathbb{N}_0$} and every \mbox{$i\in Z_1^m$}, represents
 all random variables internally generated by every scheduler, characterized by a set of parameters~\mbox{$\Xi_\ell^i=\{\theta_{\ell,j}^i\in\mathbb{R}|j\in Z_{1}^{s_i},~s_i\in\mathbb{N}\}$}, and are independent from $x_\ell^i$,~$u_\ell^i$,~$w^i_\ell$ and~$v^i_\ell$ for all \mbox{$i\in Z_1^m$} and \mbox{$\ell\in\mathbb{N}_0$}. Then any data scheduler that depends only on the primitive random variables of its corresponding control loop is said to be an admissible scheduler. Formally,
\begin{equation}\label{eq:gg}
  \delta^i_k=g(x_0^i, w_{0:k-1}^i,v_{0:k}^i, a_{0:k}^i),
\end{equation}
for $g:\mathbb{R}^{n_i}\times \mathbb{R}^{n_ik}\times\mathbb{R}^{o_i(k+1)}\times\mathbb{R}^{r_i(k+1)}\rightarrow \{0,1\}$.
\hfill\(\Box\)
\end{definition}
\par Next we introduce the concept of a tunable admissible scheduler. First, let us define the average transmission probability at every time step~\mbox{$k\in\mathbb{N}_0$} as
\begin{equation}\label{eq:averagetransmissionprobability}
p_k^i:=\mathsf{Pr}[\delta_k^i=1|\mathsf{T}_k^i],
\end{equation}
where
$\mathsf{T}_k^i:=\{\delta^i_\ell|\ell \in Z_0^{k-1}\}\cup\{\rho^i_\ell|\delta^i_\ell=1\land \ell \in Z_0^{k-1}\}.$
\begin{definition}
  \label{def:tunnableadmissiblescheduler}
(\textit{Tunable admissible scheduler}) Any admissible scheduler in~\eqref{eq:gg} is said to be a tunable admissible scheduler if its average transmission probability~\eqref{eq:averagetransmissionprobability}  at every~\mbox{$k\in\mathbb{N}_0$} is only a function of the parameters of internally generated random variables by the scheduler, i.e., $p_k^i=\phi(\Xi^i_{0:k}),$ for a given $\phi:\mathbb{R}^{s_i(k+1)}\rightarrow [0,1]$.
\hfill\(\Box\)
\end{definition}
\par Note that tunable admissible schedulers are especially convenient in the context of shared communication networks. In fact, if nodes of the network transmit stochastically with probabilities that can be controlled, one can ensure a balanced or fair use of the network. However, when the average transmission probability of the nodes are time-varying functions of the dynamic system's stochastic parameters, then its regulation is not straightforward and entails tracking the system stochastic behaviour.
\par Assume that the transmission mechanism of each user of the contention-based communication network is based on a tunable admissible scheduler with given average triggering probability \mbox{$p^i_k,~\forall i\in Z_1^m$}, at every time step. Then from a single control loop perspective, the contention-based communication network can be abstracted as if at every time step there is a probability
$q^i_k=\prod_{j=1,j\neq i}^{m}(1-p^j_k)$
that all the other users are not trying to transmit and the network is therefore available. Hence, at every time step, the control loop of interest has a successful transmission probability of
$\eta^i_k:=\mathsf{Pr}[\sigma_k^i=1|\mathsf{T}_k^i]=q^i_kp^i_k,$
which directly affects stability and the performance characteristics of that control loop, and also of the overall NCS.
\subsection{Problem statement}
In this section, we first introduce some concepts and then state the problem to be tackled in this article.  
\begin{definition}(\textit{Mean Square Stability})
We say that the system with a given scheduling and control policy is Mean Square Stable (MSS) if, for any given initial condition \mbox{$x_0\in\mathbb{R}^n$}, \mbox{$\sup\{\mathsf{E}[x^i_kx^{i\intercal}_k]\}\leqslant c$} for a \mbox{$c\in \mathbb{R}_{\geqslant 0}$} and all \mbox{$k\in\mathbb{N}_0$}. \hfill\(\Box\)
\end{definition}
\begin{definition}\label{def:pstpolicy} (\textit{Purely stochastic policy})
For any control loop of the contention-based communication network, the purely stochastic transmission (PST) policy is defined as a non-event-based tunable admissible scheduler that triggers data transmissions purely stochastically, i.e.,
    $\delta^{i,ps}_k\sim \mathsf{B}(p^i_k),$
  where \mbox{$p^i_k$} is a given triggering probability at every time  \mbox{$k\in\mathbb{N}_0$}. \hfill\(\Box\)
\end{definition}
\par The PST policy is a suitable non-event-based tunable admissible scheduler for transmitting through the contention-based networks. In fact, transmissions only depend on internal random variables parameterized by the triggering probabilities.
On the other hand, when every tunable admissible scheduler of the network uses the information set defined in~\eqref{eq:info-set-s} for deciding on data transmission, then we call the scheduling policy event-triggered or event-based.
Now, inspired by~\cite{antunes2016con}, let us define the following LQ-consistency property for any possible event-triggered control policy.
\begin{definition}(\textit{LQ-Consistent ETC})
   For any control loop, a joint event-triggered scheduling and control policy is said to be an LQ-consistent ETC policy if it results in a better performance, measured by the average quadratic cost~\eqref{eq:per}, than that of the PST policy and its associated optimal controller, while the proposed event-triggered scheduler has the same average transmission probability given in~\eqref{eq:averagetransmissionprobability} as the triggering probability of the PST policy at every time step. \hfill\(\Box\)
\end{definition}
\par Then the problem we are interested in can be stated as follows: Find an LQ-consistent decentralized ETC policy suitable for the users of the contention-based communication network, in the sense that it corresponds to a tunable admissible scheduler.
\section{Decentralized optimal control design}\label{sec:decentralizedcontrol}
\par We start by stating the following assumption.
\begin{assumption}\label{Ass:ALL_ASP}
  The schedulers of all the control loops of the shared contention-based communication network are admissible according to Definition~\ref{def:admissiblepolicy}.  \hfill\(\Box\)
\end{assumption}
\par In Theorem~\ref{th:decentralizedoptimal}, we will establish that if Assumption~\ref{Ass:ALL_ASP} holds, then the optimal control policy has a decentralized structure and follows the certainty equivalence property. In this case, one only needs to implement several locally optimal linear controllers. 
\begin{theorem}\label{th:decentralizedoptimal}
Suppose that Assumption~\ref{Ass:ALL_ASP} holds. Then for every \mbox{$i\in Z_1^m$},
  \begin{equation}\label{eq:u_optimal}
    u^i_k=K_i\mathsf{E}[x_k^i|\mathsf{H}_k^i],
  \end{equation}
   where
  \begin{equation}\label{eq:controlgain}
    \begin{aligned}
    K_i &= -(B^{\intercal}_i P_i B_i + R_i)^{-1}B^{\intercal}_i P_iA_i, \\
    P_i & = A^{\intercal}_i P_i A_i + Q_i-K^{\intercal}_i (B^{\intercal}_i P_i B_i + R_i)K_i,
\end{aligned}
  \end{equation}
    is an optimal control policy in the sense that it minimizes the average quadratic cost~$J^i$ given in~\eqref{eq:per}.
\hfill\(\Box\)
\end{theorem}
\par The proof procedure is similar to the one presented in \cite[Lemma 2.1]{Moin2014Thesis}. The full details can be found in \cite{balaghiinaloo2019decentralized}.
\section{Purely stochastic transmission policy}\label{sec:Purelystochastic}
In this section, we analyze a given control loop of the shared contention-based communication network, when its scheduler is operating based on the PST policy introduced in Definition~\ref{def:pstpolicy}. First, we state that, given Assumption~\ref{Ass:ALL_ASP}, the remote state estimation of this control loop is determined by a linear iteration, which follows from the results in~\cite{Imer2006optimal,schenato2007foundations}. Second, the closed-form expression of the local average quadratic cost~\eqref{eq:per} is derived when all the schedulers have constant triggering probabilities at every time step.
\begin{corollary}\label{lem:PSTestimator}
(\cite{Imer2006optimal,schenato2007foundations})
Suppose that Assumption~\ref{Ass:ALL_ASP} holds and consider a given control loop with scheduler operating based on the PST policy according to Definition~\ref{def:pstpolicy}, the remote state estimation needed for the calculation of the optimal control policy in~\eqref{eq:u_optimal} is given by \mbox{$\bar{x}^i_{k|k}=\mathsf{E}[x^i_k|\mathsf{H}^i_k]$}, where at every~\mbox{$k\in\mathbb{N}_0$},
\begin{equation}\label{eq:Observer}
\begin{aligned}
\bar{x}^i_{k+1|k} &= A_i\bar{x}^i_{k|k}+B_iu^i_{k},~~~~
\bar{x}^i_{k|k} =\begin{cases}
                    \hat{x}^i_{k|k}, & \mbox{if } \sigma^i_k=1 \\
                    \bar{x}^i_{k|k-1}, & \mbox{otherwise}.
                  \end{cases}
\end{aligned}
\end{equation}
\hfill\(\Box\)
\end{corollary}
\par The MSS and the performance of the control loops depend on the data transmission policies. The following lemma provides a sufficient condition for MSS and the optimal control performance of each subsystem, when its scheduler is operating based on the PST policy.
\begin{lemma}\label{lem:PSTcost}
Consider that Assumption~\ref{Ass:ALL_ASP} holds and all the schedulers are transmitting with constant average probabilities at every time step, i.e.,
\mbox{$p^i_k=p^i,~\forall i\in Z_1^m$}, and at all \mbox{$k\in\mathbb{N}_0$}. Then the control loop~\mbox{$i\in Z_1^m$} with the PST scheduling policy and the optimal controller~\eqref{eq:u_optimal} characterized by~\eqref{eq:controlgain} and~\eqref{eq:Observer} is MSS if,
\begin{equation}\label{eq:stabilitycondition}
\varrho(\sqrt{1-q^ip^i}A_i)<1,
\end{equation}
where \mbox{$q^i=\prod_{j=1,j\neq i}^{m}(1-p^j)$}. Moreover, if~\eqref{eq:stabilitycondition} holds, the optimal average quadratic cost of the subsystem is
\begin{equation}\label{eq:pscost}
\begin{aligned}
J^i_{ps}=\text{tr}(P_iW_i)+&\sum_{j=0}^{\infty}[(1-q^i p^i)^{j+1}\text{tr}(A^j_iW_iA^{j\intercal}_i Y_i)&\\&
+q^ip^i(1-q^i p^i)^{j}\text{tr}(A^j_i\Theta_i A^{j\intercal }_i Y_i)].
\end{aligned}
\end{equation}
where \mbox{$Y_i=K^{\intercal}_i (B^{\intercal}_i P_i B_i + R_i)K_i$}.
\hfill\(\Box\)
\end{lemma}
\par The first part of Lemma~\ref{lem:PSTcost} is based on the results of~\cite{Imer2006optimal,schenato2007foundations}, however, the full proof can be found in~\cite{balaghiinaloo2019decentralized}.
\par A key question is how to find the optimal values of the triggering probabilities in order to minimize a social performance index. If the scheduler of every control loop using the network follows the PST policy with a constant success rate at every time step, then a natural answer is to consider the following optimization problem
\begin{equation}\label{eq:totalperformance}
  (p^1,\dots, p^m)^*=\arg\min_{p^1,\dots,p^m}\sum_{i=1}^{m}J^i_{ps}.
\end{equation}
However, \eqref{eq:totalperformance} is in general a non-convex and non-separable problem, and therefore, it is not easily solvable. In Section~\ref{sec:regulation}, we determine the constant triggering probabilities of the schedulers based on a network utility maximization criterion, which is more tractable.
\section{Proposed scheduling policy using stochastic thresholds}\label{sec:proposedpolicy}
\par The novel tunable admissible scheduler proposed in this work combines the features of the STETT~\cite{Brunner2018stochastic} and the PST policies. In this section, we first introduce the STETT policy and discuss its advantages. Then, we propose a combined scheduling policy which is a tunable admissible scheduler.
\subsection{Stochastic threshold event-triggered transmission}
\par Inspired by~\cite{Brunner2018stochastic}, the data triggering mechanism of the STETT policy for every linear system with Gaussian disturbance and measurement noise is defined as
\begin{equation}\label{eq:scheduler2}
\delta^{i,st}_k:=\begin{cases}
             1, & \mbox{if } \mbox{$\frac{1}{2}e^{i\intercal}_{k|k-1}\Psi_{k|k-1}^{i,-1}e^i_{k|k-1}>r^i_k$}, \\
             0, & \mbox{otherwise},
           \end{cases}
\end{equation}
in which \mbox{$r^i_k\sim \mathsf{exp}(\lambda^i_k)$} for \mbox{$\lambda^i_k\in\mathbb{R}_{\geqslant 0}$} is exponentially distributed random threshold, \mbox{$e^i_{k|k-1}:=\hat{x}^i_{k|k}-\bar{x}^i_{k|k-1}$} and \mbox{$e^i_{k|k}:=\hat{x}^i_{k|k}-\bar{x}^i_{k|k}$} are the predicted and the updated errors between the local and the remote state estimations with the covariances \mbox{$\Psi^i_{k|k-1}:=\mathsf{E}[e^i_{k|k-1}e^{i\intercal}_{k|k-1}|\mathsf{T}^i_k]$} and \mbox{$\Psi^i_{k|k}:=\mathsf{E}[e^i_{k|k}e^{i\intercal}_{k|k}|\mathsf{T}^i_{k+1}]$}, respectively. At every time step, the value of \mbox{$e^i_{k|k-1}$} can be determined by subtracting the updated state estimation by the Kalman filter~\eqref{eq:scheduler-estimate} ($\hat{x}_{k|k}^i$) and the predicted state estimation ($\bar{x}^i_{k|k-1}$) determined based on~\eqref{eq:Observer}.  
\par Unlike deterministic threshold event-triggered transmission policies, the STETT policy ensures that \mbox{$e^i_{k|k-1}$} remains Gaussian distributed at all time steps when there is no data collision or drop-out \cite{han2017stochastic}.
However, as we shall see shortly, although after a successful transmission, $e^i_{k|k-1}$ is Gaussian distributed, and remains to be so until the first new attempt to transmit, in case of data collision in the current transmission attempt, the distribution of $e^i_{k|k-1}$ will become the sum of two Gaussians at the following time step.
\par To clarify this statement, note that the error used in the scheduling law~\eqref{eq:scheduler2} has the following dynamics
\begin{equation}\label{eq:schedulerserror}
  e^i_{k+1|k}=(1-\sigma^i_k)A_ie^i_{k|k}+L_i(C_i\hat{e}^i_{k+1|k}+v^i_{k+1}),
\end{equation}
 where \mbox{$\hat{e}^i_{k+1|k}=x^i_k-\hat{x}^i_{k|k-1}$} is the state estimation error of the Kalman filter which has a Gaussian distribution with zero mean and \mbox{$\bar{\Theta}_i$} as its covariance matrix at all time steps. Note that the error dynamics in~\eqref{eq:schedulerserror} depends only on the information set
 $\mathsf{R}^i_k=\{v^i_{\ell},w^i_{\ell},r^i_{\ell},\rho^i_{\ell}|{\ell}\in Z^{k-1}_0\}\cup\{v^i_{k}\} \cup\{x^i_0\}$ at every $k\in\mathbb{N}_0$,
 which indicates that the STETT policy~\eqref{eq:scheduler2} is a function of the primitive random variables, and therefore, it is as an admissible scheduler according to Definition~\ref{def:admissiblepolicy}. 
 \par Now if we let \mbox{$\sigma^i_{k-1}=1$}, then \mbox{$e^i_{k|k-1}=L_i(C_i\hat{e}^i_{k|k-1}+v^i_{k})$}, which is clearly Gaussian. Assume now that the distribution of the predicted error ($e^i_{k|k-1}$) is Gaussian with an arbitrary covariance \mbox{$\Psi^i_{k|k-1}$} at a time step \mbox{$k\in\mathbb{N}_0$}, i.e.,
\begin{equation}\label{eq:Gaussian}
  \mathsf{f}(e^i_{k|k-1}|\mathsf{T}^i_{k})=\mathsf{N}(0,\Psi^i_{k|k-1}).
\end{equation}
The next lemma shows that the pdf of the predicted error at the next time step, i.e., \mbox{$k+1$}, is Gaussian in case of no data triggering $(\delta^{i,st}_k=0)$, while in case of a data triggering and collision \mbox{$(\delta^{i,st}_k=1\land \rho^i_k=0)$} the pdf of the predicted error becomes the sum of two Gaussians.
\begin{lemma}\label{le:twogaussians}
Assume that the distribution of the predicted error follows \eqref{eq:Gaussian} at a time step \mbox{$k\in\mathbb{N}_0$}. Then
\begin{equation}\label{eq:schedulerprob1}
p^i_k=\mathsf{Pr}(\delta^{i,st}_{k}=1|\mathsf{T}^i_{k})=1-(1+\lambda^i_k)^{-\frac{n_i}{2}},
\end{equation}
is the average data transmission probability by the STETT scheduler \eqref{eq:scheduler2} at time step $k$. Moreover,
\begin{equation}\label{eq:nottriggeringpdf}
\mathsf{f}(e^i_{k+1|k}|\delta^{i,st}_{k}=0, \mathsf{T}^i_{k})=\mathsf{N}(0,\hat{\Psi}^i_{k+1|k}),
\end{equation}
and
\begin{equation}\label{eq:collisionpdf}
\begin{aligned}
\mathsf{f}(&e^i_{k+1|k}|\delta^{i,st}_k=1,\rho^i_k=0, \mathsf{T}^i_{k})&\\&=\frac{1}{p^i_k}\mathsf{N}(0,\Psi^i_{k+1|k})-\frac{1-p^i_k}{p^i_k}\mathsf{N}(0,\hat{\Psi}^i_{k+1|k}),
\end{aligned}
\end{equation}
are the pdfs of the predicted error at time step \mbox{$k+1$} in case of no data triggering \mbox{$(\delta^{i,st}_k=0)$}
 and data collision \mbox{$(\delta^{i,st}_k=1\land \rho^i_k=0)$}, respectively, where
\begin{equation}\label{eq:updatedcovariance}
    \begin{aligned}
\Psi^i_{k+1|k}&=A\Psi^i_{k|k-1}A_i^{\intercal}+\Phi_i,&\\
\hat{\Psi}^i_{k+1|k}&=\frac{1}{1+\lambda^i_k}A_i\Psi^i_{k|k-1}A_i^{\intercal}+\Phi_i,
\end{aligned}
\end{equation}
in which $\Phi_i=A_i\Theta_i A_i^{\intercal}-\Theta_i +W_i$.\hfill\(\Box\)
\end{lemma}
\par The proof follows by applying the Bayes law of conditional probability. It is omitted due to space restrictions and can be found in \cite{balaghiinaloo2019decentralized}.
\par From~\eqref{eq:schedulerprob1}  we can conclude that in a time period between the last successful transmission and the first subsequent data collision the triggering probability depends only on threshold parameter \mbox{$\lambda_k^i$}. Therefore, we can easily regulate it as follows
\begin{equation}\label{eq:lambda}
  \lambda^i_k=(1-p^i_k)^{-\frac{2}{n_i}}-1.
\end{equation}
However, following similar arguments to the ones just given, it can be shown that in between every two consecutive successful transmissions, every collision doubles the number of Gaussian terms of the predicted state estimation error pdf \cite{balaghi2018optimal}. When the number of Gaussian terms is more than one, the triggering probability depends not only on the parameter of the random threshold ($\lambda_k^i$) but also on the covariances of the multiple Gaussian terms. Therefore, it is not trivial to regulate the triggering probability desirably after the first collision instance. This motivates the next scheduling policy.
\subsection{Combined event-triggered transmission policy}
\par In this section, we propose a combined event-triggered transmission (CETT) policy \mbox{$\pi^i = (\mu^i_0,\mu^i_1,\mu^i_2,\dots)$}, where \mbox{$\mu^i_k: \mathsf{L}^i_k\rightarrow \{0,1\}$} and \mbox{$\delta^{i,\mu}_k=\mu^i_k(\mathsf{L}^i_k)$}, which inherits the advantages of the STETT policy and is in the class of tunable admissible schedulers. Based on this policy, after every successful transmission, the scheduler triggers based on the STETT policy with any desired probability $p^i_k$ up to the time step at which the first collision happens. After that, the scheduler keeps triggering based on the PST policy with the desired probability $p^i_k$ until the next successful transmission time. This process is repeated between every two successive successful transmissions.
\begin{definition}\label{def:proposedscheduling}(\textit{CETT})
    Let \mbox{$\bar{\ell}^i_k:=\max\{\ell< k|\sigma^i_\ell=1\}$} be the time of the last successful transmission before the current time step $k$. Then, we can specify the CETT policy \mbox{$\delta^{i,\mu}_k$} as follows
\begin{equation}\label{eq:scheduler1}
\delta^{i,\mu}_k=\begin{cases}
\delta^{i,st}_k, & \mbox{if } \mbox{$k\!=\!\bar{\ell}^i_{k}+1$} \mbox{ or } \mbox{$(\delta^{i,st}_{\!\bar{\ell}^i_{k}+1},\!\dots\!,\delta^{i,st}_{k-1}) \!= \!(0,\dots,0)$}\\
\delta^{i,ps}_k, & \mbox{otherwise},
\end{cases}
\end{equation}
where $\delta^{i,st}_k$ follows \eqref{eq:scheduler2} with $\lambda^i_k$ determined by \eqref{eq:lambda} for a given $p^i_k$ as the average triggering probability and  \mbox{$\delta^{i,ps}_k\sim\mathsf{B}(p^i_k)$} for all corresponding time steps.
\hfill\(\Box\)
\end{definition}
\par We know that the STETT policy is only easily tunable after every successful data transmission time up to the next data triggering time. As a result, the CETT policy in~\eqref{eq:scheduler1} follows the STETT policy as long as it is easily tunable and we can use~\eqref{eq:lambda} to regulate its triggering probability at these time steps. However, at every transmission epoch, the CETT policy starts following the PST policy right after the first data collision time, which can also be easily tuned.
\section{Main results}\label{sec:mainresults}
\par In this section, we state the main results of the article.
We start by introducing the optimal control policy associated with the CETT policy~\eqref{eq:scheduler1} in Theorem~\ref{th:opt}.

\begin{theorem}\label{th:opt}
Suppose that Assumption~\ref{Ass:ALL_ASP} holds and for a given control loop the scheduler follows the CETT policy with a given set of average triggering probabilities \mbox{$\textbf{P}^i=\{p^i_k|k\in\mathbb{N}_0\}$}. Then the control policy \eqref{eq:u_optimal}, characterized by \eqref{eq:controlgain} and \eqref{eq:Observer}, is optimal in the sense that it minimizes its corresponding average quadratic cost given in~\eqref{eq:per}. \hfill\(\Box\)
\end{theorem}
\par The proof of Theorem~\ref{th:opt} can be found in \cite{balaghiinaloo2019decentralized}, in which we show that the pdf of the remote state estimation associated with the CETT policy~\eqref{eq:scheduler1} follows a sum of Gaussians, and at every time step, all Gaussian terms have the same mean values determined according to~\eqref{eq:Observer}. 
\begin{definition}\label{def:CETC}(\textit{CETC})
The combination of the CETT policy~\eqref{eq:scheduler1} and its corresponding optimal control law~\eqref{eq:u_optimal},~\eqref{eq:controlgain} and~\eqref{eq:Observer} is denoted by the CETC policy. \hfill\(\Box\)
\end{definition}
\par The next theorem states the main result of the article establishing the LQ-consistency of the CETC policy.
\begin{theorem}\label{th:consistency}
Suppose that Assumption~\ref{Ass:ALL_ASP} holds and consider a control loop when its scheduler follows the PST policy for a given set of triggering probabilities \mbox{$\textbf{P}^i=\{p^i_k|k\in\mathbb{N}_0\}$}. Suppose that this control loop is MSS when its controller follows the optimal policy \eqref{eq:u_optimal} characterized by \eqref{eq:controlgain} and \eqref{eq:Observer}. Then, the average quadratic cost~\eqref{eq:per} of this control loop  when its scheduler-controller is operating based on the CETC policy is strictly smaller than that of the optimal control performance when the scheduler is operating based on the PST policy with identical set of triggering probabilities $\textbf{P}^i$, i.e.,
	$ J^i_\pi<J^i_{ps}.$\hfill\(\Box\)
\end{theorem}
\par The proof of Theorem \ref{th:consistency} is available in the Appendix.
\begin{remark}\label{rem:4}
  Based on Theorem \ref{th:consistency}, the MSS of each control loop with the scheduler following the PST policy for a given set of triggering probabilities \mbox{$\textbf{P}^i=\{p^i_k|k\in\mathbb{N}_0\}$} and its associated optimal controller can also guarantee the MSS when the control loop is operating based on the CETC policy with identical set of triggering probabilities $\textbf{P}^i$. As an example, when \mbox{$p^i_k=p^i,~\forall i\in Z_1^m,~\forall k\in \mathbb{N}_0$}, then the control loops operating based on the CETC policy are MSS if~\eqref{eq:stabilitycondition} holds.
\end{remark}
\section{Decentralized implementation and network utility maximization}\label{sec:regulation}
\par From the discussions so far in this article, it can be concluded that all the schedulers and the control policies only require local information and, therefore, can be implemented in a decentralized fashion. In this section, we discuss how to regulate the triggering probability of the schedulers decentrally in order to optimize a social criterion.
\par We take a notion of network utility from~\cite{lee2006jointly}, which considers a weighted proportional fairness between the network users. Based on that, we assume a constant triggering probability at every time step for every user (therefore, we drop its time index $k$ for simplicity) and assign to it a network utility allocation function as follows
 $ U^i(\eta^i)=c^i\log (\eta^i),$
where \mbox{$\eta^i=p^i\prod_{j=1,j\neq i}^{m}(1-p^j)$} is the constant successful transmission probability and the constant \mbox{$c^i\in\mathbb{R}_{>0}$} determines the transmission priority of every user, which can be selected based on the average quadratic performance of every control loop~\eqref{eq:pscost}. Then the optimal successful transmission probabilities are determined as follows
\begin{equation}\label{eq:etaoptimization}
  (\eta^{1*},\dots,\eta^{m*})=\arg\max_{\eta^1\dots \eta^m} \sum_{j=1}^{m}U^j(\eta^j).
\end{equation}
The above optimization problem results in \mbox{$p^{i*}=c^i/\sum_{j=1}^{m}c^j$} as the optimal triggering probability for every scheduler (\mbox{$i\in Z_1^m$}) of the considered contention-based communication network.
Now we suggest the following methods for the selection of $c^i$ for every \mbox{$i\in Z_1^m$}:
\begin{enumerate}[label=(\roman*)]
  \item The coefficient \mbox{$c^i$} for every subsystem \mbox{$i\in Z_1^m$} can be selected based on the parameters affecting its average quadratic performance. According to Lemma~\ref{lem:PSTcost}, a natural choice is
 $c^i=\alpha_i\text{tr}(A_iW_iA_i^{\intercal } Y_i)+(1-\alpha_i)\text{tr}(A_i\Theta_i A_i^{\intercal} Y_i)$
 where \mbox{$\alpha_i\in [0,1]$} can be selected arbitrarily.
  \item We can also select all transmission priorities to be equal which results in \mbox{$p^{i*}=1/m$} for all \mbox{$i\in Z_1^m$}. This triggering probability is equal to the optimal triggering probability for maximizing the throughput of the slotted-ALOHA communication channel \cite{kurose2005computer}.
\end{enumerate}
Both strategies for tuning the triggering probabilities can be implemented in a decentralized fashion, as long as every node in the network has access to \mbox{$c^i$} of all nodes. 
\section{Numerical example}\label{sec:Numericalresult}
\begin{figure}[!t]
    \centering
    \includegraphics[width=8cm]{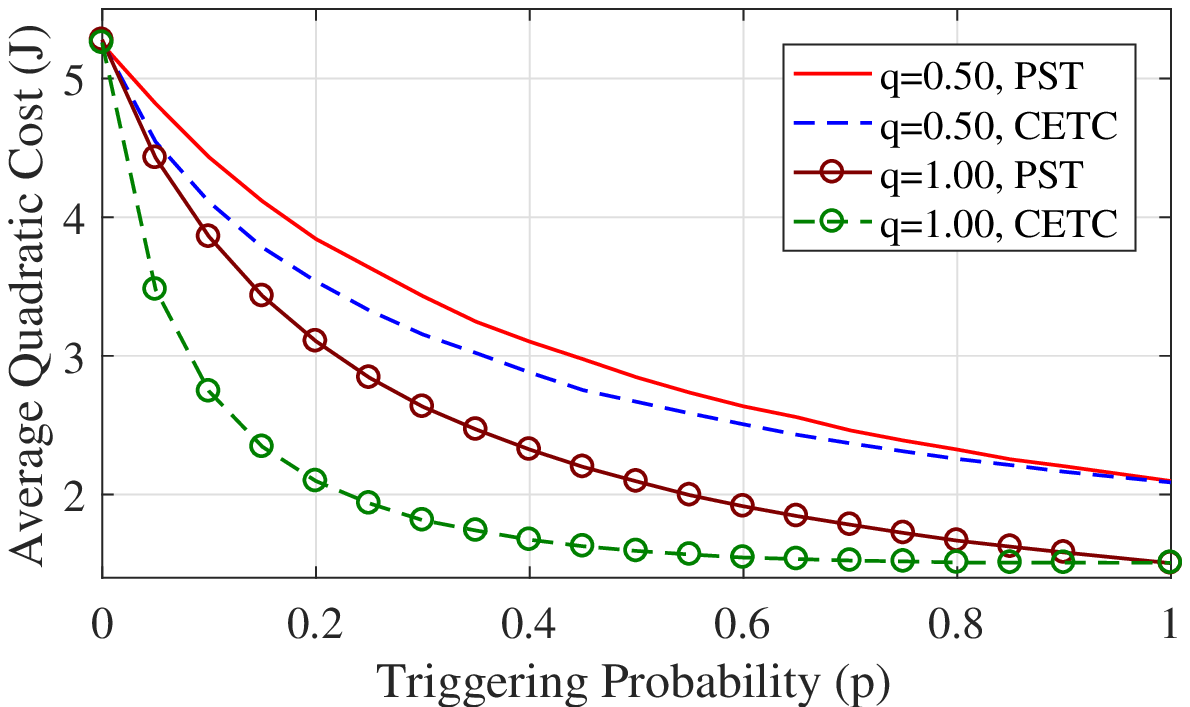}
    \caption{Average quadratic performance comparison between the PST and the CETC policies.}
    \label{fig:cost}
    \centering
    \includegraphics[width=8cm]{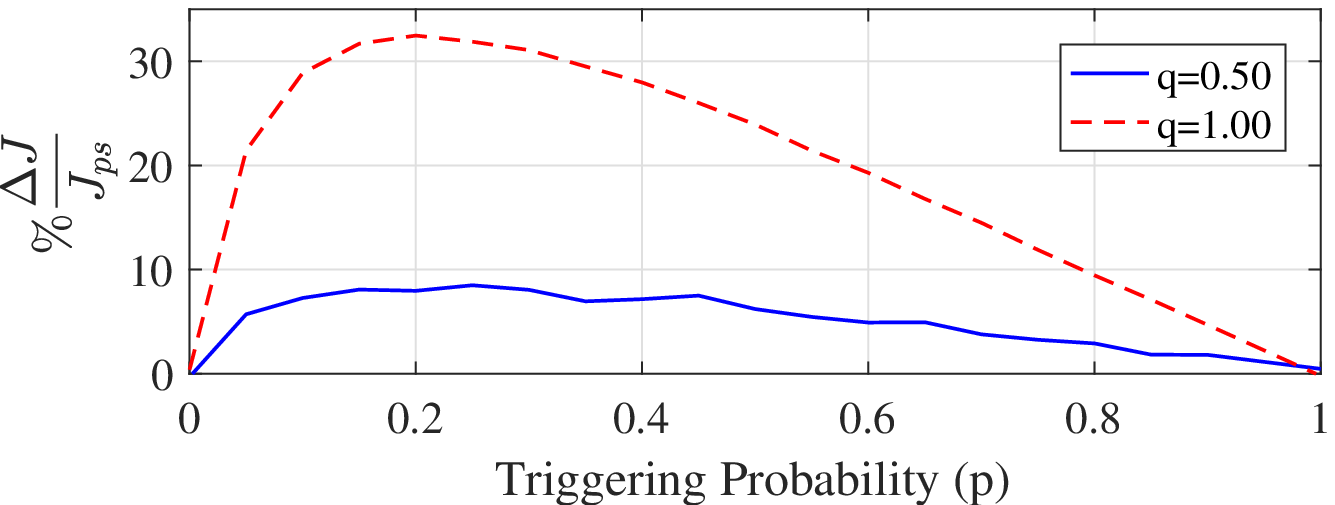}
    \caption{Average quadratic performance gain percentage of the CETC policy in comparison with the PST policy.}
    \label{fig:dcost}
\end{figure}
In this section, we illustrate via a numerical example that the proposed CETC policy outperforms the PST policy and we assert the performance gains. Consider a scalar LTI subsystem with \mbox{$A=0.9$}, \mbox{$B=1$}, \mbox{$C=1.5$}, \mbox{$W=1$}, and \mbox{$V=1.5$}. Due to the decentralized structure of the proposed policies, we can consider just a single control loop of this NCS and drop the index $i$ of the control loop parameters. Moreover, let \mbox{$Q\!=\!1$} and \mbox{$R\!=\!0.1$} be the parameters of the average quadratic performance. Then, the state feedback controller and the Kalman filter gains are \mbox{$K=-0.8233$} and \mbox{$L=0.4476$}, respectively. We consider a constant triggering probability for the scheduler of this control loop at all time steps. In Fig.~\ref{fig:cost}, we compare the average quadratic performance of both policies for two different constant probabilities that the network is free \mbox{$q\in\{0.50,1.00\}$} for this control loop at all time steps. These plots illustrate what we have observed in Monte-Carlo simulations. For each pair of \mbox{$(p,q)$}, we consider \mbox{$n_{MC}=10$} as the number of Monte-Carlo runs, where for each of them, \mbox{$T=10^5$} is the total number of simulation time steps. The initial state for all simulations is assumed to be zero, i.e. \mbox{$x_0=0$}. Fig.~\ref{fig:dcost} shows the percentage of the performance gains of the CETC policy with respect to the purely stochastic policy, i.e., \mbox{$\%\Delta J/J_{ps}$}, where~\mbox{$\Delta J=J_{ps}-J_{\pi}$}. As can be seen, when the availability probability of the network $(q)$ for a specific control loop is higher, the performance gain obtained by the CETC policy is also higher.
\section{Conclusion}\label{sec:conclusion}
This work considers multiple independent linear systems communicating through a shared contention-based communication network with their local remote controllers. We introduce a class of admissible schedulers which provides a decoupled optimal control design structure for the users of the shared contention-based communication network and is proved to have the certainty equivalence property. Then, two scheduling policies in this class of admissible schedulers are introduced, a non-event-based and an event-based policy. Moreover, their triggering probabilities are easily tunable at every time step. This feature can be used for maximizing the utility of the network in the sense of providing proportional fairness between the users of the network. It is proved that for both of these scheduling policies, the local optimal control law is determined based on a Kalman filter state estimator. The main contribution of this article is the LQ-consistency of the proposed event-based control strategy, i.e., for any subsystem, the loop with the event-based scheduler and its optimal control law outperforms the non-event-based scheduling policy with its associated optimal control law, as the triggering probability of both scheduling policies are the same at every time step.
\appendix
\setcounter{equation}{0}
\renewcommand{\theequation}{\thesection.\arabic{equation}}
\subsection{Proof of Theorem \ref{th:consistency}}
\par Let us consider a single control loop and drop the index~$i$ for simplicity. Consider \mbox{$N\geqslant 1$} as the period for every two successive successful transmissions, then
\begin{equation}\label{eq:per2}
\begin{aligned}
J_a&= \text{tr}(PW)+\frac{1}{\mathsf{E}[N]}\mathsf{E}[\sum_{t=0}^{N-1}\text{tr}(Y\Gamma^a_{t|t})]&\\&=
\text{tr}(PW)+\sum_{v=1}^{\infty} \frac{\mathsf{Pr}(N=v)}{\mathsf{E}[N]} \sum_{t=0}^{v-1}\text{tr}(Y\Gamma^a_{t|t}),
\end{aligned}
\end{equation}
which can be established using Wald's identity as in~\cite{Brunner2018stochastic}, where
\mbox{$\Gamma^a_{t|t}=\mathsf{E}[\bar{e}_{t|t}\bar{e}^\intercal_{t|t}|\hat{x}_{0|0}, \mathsf{I}_t]$} for \mbox{$\mathsf{I}_t=\{\sigma_k=0|k\in Z_1^t\}$}, \mbox{$\bar{e}_{t|t}=x_t-\bar{x}_{t|t}$}, for all \mbox{$t\in\mathbb{N}_0$} and \mbox{$ a\in\{ps,\mu\}$}. Given the assumption that both scheduling policies trigger with the same probabilities at every time, to prove Theorem~\ref{th:consistency}, it is sufficient to prove
\begin{equation}\label{eq:costinequality2}
  \hat{J}_{\mu}^v<\hat{J}_{ps}^v,
\end{equation}
where \mbox{$\hat{J}_a^v=\sum_{t=1}^{v-1}\text{tr}(Y\Gamma^a_{t|t})$} for \mbox{$ a\in\{ps,\mu\}$} and all \mbox{$v\in\mathbb{N}_{\geqslant 2}$} (at \mbox{$t=0$} both policies result in the same cost values). We shall proof~\eqref{eq:costinequality2} by using induction on $v$. Suppose \mbox{$v=2$}, then for every \mbox{$a\in\{\mu,ps\}$}, we have
$\hat{J}_a^2=\text{tr}(Y\Gamma^a_{1|1})=\sum_{i=1}^{3}\text{tr}\big(Y\bar{\Gamma}^a_{1|1}(i)\big)S^a(i),$
 where for every \mbox{$i\in \{1,2,3\}$},
$$
\begin{aligned}
\bar{\Gamma}^a_{1|1}(i)=&\mathsf{E}[\bar{e}_{1|1}\bar{e}^\intercal_{1|1}|\mathsf{v}^a(i)],~S^a(i)=\mathsf{Pr}(\delta^a_1=m(i),\rho_1=l(i)),
\end{aligned}
$$
for
\begin{equation}\label{eq:INFO}
   \mathsf{v}^a(i)=\{\hat{x}_{0|0},\delta_1^a=m(i),\rho_1=l(i)\},
 \end{equation}
in which
\begin{equation}\label{eq:ml2}
    \big(m(i),l(i)\big)=\begin{cases}
              (0,0), & \mbox{if } i=1 \\
              (1,0), & \mbox{if } i=2 \\
              (0,1), & \mbox{if } i=3.
            \end{cases}
  \end{equation}
In Table~\ref{tab:table11}, we determine the values of~\mbox{$\bar{\Gamma}^a_{1|1}(i)$} and \mbox{$S^a(i)$}. Consider \mbox{$r_0:=\text{tr}(Y\Theta)$}, \mbox{$r_1:=\text{tr}(Y\Phi)$} for \mbox{$\Phi=A\Theta A^\intercal-\Theta+W$}, \mbox{$r_2:=\frac{r_1}{1+\lambda_1}$} and \mbox{$r_3:=\frac{1}{p_1}r_1-\frac{1-p_1}{p_1}r_2$}, which are used in the tables.
  \begin{table}[t!]
  \begin{center}
    \caption{Performance terms of the CETC and the PST policies, when \mbox{$v=2$}.}
    \label{tab:table11}
    \begin{tabular}{l V{2} c | c|c|c} 
      $i$ & $\delta^\mu_1$ & $\rho_1$ &  $S^{\mu}(i)$ & $\text{tr}(Y\bar{\Gamma}^\mu_{1|1}(i))$ \\
      \hline
     1&  0  &   0    & \mbox{$(1-q_1)(1-p_1)$}& $r_0+r_{2}$        \\
     2&  1  &   0    & \mbox{$(1-q_1)p_1$}&    $r_0+r_{3}$       \\
     3&  0  &   1    &  \mbox{$q_1(1-p_1)$}&             $r_0+r_{2}$     \\
    \end{tabular}
  \end{center}
  \begin{center}
    \label{tab:table12}
    \begin{tabular}{l V{2} c |c | c|c} 
      $i$ & $\delta^{ps}_1$ & $\rho_1$   & $S^{ps}(i)$ &  $\text{tr}(Y\bar{\Gamma}^{ps}_{1|1}(i))$ \\
      \hline
     1&  0 \text{or} 1  &    0  & \mbox{$(1-q_1)$}&    $r_0+r_{1}$      \\
     2&  0  &    1 &  \mbox{$q_1(1-p_1)$}&                $r_0+r_{1}$      \\
    \end{tabular}
  \end{center}
\end{table}
Now based on the table we have
$$
\begin{aligned}
&\hat{J}_\mu^2=(1-q_1)(r_0+r_1)+q_1(1-p_1)(r_0+r_2),&\\&
\hat{J}_{ps}^2=(1-q_1)(r_0+r_1)+q_1(1-p_1)(r_0+r_1),
\end{aligned}
$$
which indicates that \mbox{$\hat{J}_\mu^{2}< \hat{J}_{ps}^{2}$} since \mbox{$r_2<r_1$}. Then by assuming that \eqref{eq:costinequality2} holds for \mbox{$v=z$}, i.e.,
\mbox{$\hat{J}_{\mu}^{z}<\hat{J}_{ps}^{z}$},
 we should prove the same inequality for \mbox{$v=z+1$}. For this purpose, we need the next proposition, whose proof can be found in~\cite{balaghiinaloo2019decentralized}.
\begin{proposition}\label{Prop:11}
Consider $\Psi^a_{1|1}(i)=\mathsf{E}[{e}_{1|1}{e}^\intercal_{1|1}|\mathsf{v}^a(i)]$ as the updated covariance of the error at \mbox{$t=1$} for every transmission epoch with \mbox{$v> 2$}, where \mbox{$i\in\{1,2,3\}$} and \mbox{$\mathsf{v}^a(i)$} is characterized by~\eqref{eq:INFO} and~\eqref{eq:ml2}. Then for both \mbox{$a\in\{\mu,ps\}$}, 
$$
\begin{aligned}
\sum_{t=2}^{v}\text{tr}(Y\Gamma^a_{t|t})&|\mathsf{v}^a(i)=\text{tr}\big(L_a(\Psi_{1|1}^a(i)+\Theta)\big)+\hat{J}_a^{v},
\end{aligned}
$$
where \mbox{$L_a$} is a positive definite matrix such that \mbox{$L_\mu<L_{ps}$}, and
\mbox{$\Psi^a_{1|1}(i):=\mathsf{E}[e_{1|1}e^{\intercal}_{1|1}| \mathsf{v}^a(i)]$}.\hfill\(\Box\)
\end{proposition}
\par For every \mbox{$ a\in\{ps,\mu\}$}, we have
$$\hat{J}_a^{z+1}=\sum_{i=1}^{3}\Big(\text{tr}\big(Y\bar{\Gamma}^a_{1|1}(i)\big)+\sum_{t=2}^{z}\text{tr}(Y\Gamma^a_{t|t})|\mathsf{v}(i)\Big)S^a(i),$$
and based on Proposition~\ref{Prop:11}, it is simplified as
$$\hat{J}_a^{z+1}=\sum_{i=1}^{3}\big(\text{tr}(Y\bar{\Gamma}^a_{1|1}(i))+\text{tr}\big(L_a(\Psi^a_{1|1}(i)+\Theta)\big)+\hat{J}_a^{z}\big)S^a(i).$$
Therefore, by considering the assumption of induction, in order to prove \mbox{$\hat{J}_{\mu}^{z+1}<\hat{J}_{ps}^{z+1}$}, we just need to prove
\mbox{$\hat{j}_{\mu}^{z+1}<\hat{j}_{ps}^{z+1}$},
where for every \mbox{$a\in\{\mu,ps\}$},
\mbox{$\hat{j}_a^{z+1}=\sum_{i=1}^{3}C^a(i)S^a(i)$},
in which
\mbox{$C^a(i)=\text{tr}\big(Y\bar{\Gamma}^a_{1|1}(i)\big)+\text{tr}\big(L_a\Psi^a_{1|1}(i)\big)$}.
Then let us denote \mbox{$s_1:=\text{tr}(L_{ps}\Phi)$}, \mbox{$s_2:=\text{tr}(L_{\mu}\Phi)/(1+\lambda_1)$}, \mbox{$s_3:=\frac{1}{p_1}(1-\frac{1-p_1}{1+\lambda_1})\text{tr}(L_{\mu}\Phi)$} and  \mbox{$s_4:=\text{tr}(L_{\mu}\Phi)$}.
Based on Proposition~\ref{Prop:11},
\begin{equation}\label{eq:s11}
s_2<s_4<s_1.
\end{equation}
In Table~\ref{tab:table3} the values of \mbox{$C^a(i)$} are given for the CETC and the PST policies, respectively. Then we have
$$
\begin{aligned}
&\hat{j}_\mu^{z+1}=(1-q_1)(r_0+r_1+s_4)+q_1(1-p_1)(r_0+r_2+s_2),&\\&
\hat{j}_{ps}^{z+1}=(1-q_1)(r_0+r_1+s_1)+q_1(1-p_1)(r_0+r_1+s_1),
\end{aligned}
$$
and by using the inequalities given in~\eqref{eq:s11}~and \mbox{$r_2<r_1$}, we can infer
\mbox{$\hat{j}_\mu^{z+1}< \hat{j}_{ps}^{z+1}$},
which completes the proof.
  \begin{table}[h!]
  \begin{center}
    \caption{The \mbox{$C^a(i)$} values for the CETC and the PST policies.}
    \label{tab:table3}
    \begin{tabular}{l V{2} c | c|c|c} 
      $i$ &  $\delta^\mu_1$ & $\rho_1$ &  $S^{\mu}(i)$ & $C^\mu(i)$ \\
      \hline
     1&  0  &   0    & \mbox{$(1-q_1)(1-p_1)$}& $r_0+r_2+s_{2}$        \\
     2&  1  &   0    & \mbox{$(1-q_1)p_1$}&    $r_0+r_3+s_{3}$       \\
     3&  0  &   1    &  \mbox{$q_1(1-p_1)$}&             $r_0+r_2+ s_{2}$     \\
    \end{tabular}
  \end{center}

  \begin{center}
    \begin{tabular}{l V{2} c |c | c|c} 
      $i$   & $\delta^{ps}_1$ & $\rho_1$ & $S^{ps}(i)$ &  $C^{ps}(i)$ \\
      \hline
     1&  0 \text{or} 1  &    0 & \mbox{$(1-q_1)$}&    $r_0+r_1+s_{1}$      \\
     2&  0  &    1 &  \mbox{$q_1(1-p_1)$}&                $r_0+r_1+s_{1}$      \\
    \end{tabular}
  \end{center}
\end{table}
\bibliographystyle{IEEEtran}
\bibliography{thesis}

\begin{thebibliography}{10}
\providecommand{\url}[1]{#1}
\csname url@samestyle\endcsname
\providecommand{\newblock}{\relax}
\providecommand{\bibinfo}[2]{#2}
\providecommand{\BIBentrySTDinterwordspacing}{\spaceskip=0pt\relax}
\providecommand{\BIBentryALTinterwordstretchfactor}{4}
\providecommand{\BIBentryALTinterwordspacing}{\spaceskip=\fontdimen2\font plus
\BIBentryALTinterwordstretchfactor\fontdimen3\font minus
  \fontdimen4\font\relax}
\providecommand{\BIBforeignlanguage}[2]{{%
\expandafter\ifx\csname l@#1\endcsname\relax
\typeout{** WARNING: IEEEtran.bst: No hyphenation pattern has been}%
\typeout{** loaded for the language `#1'. Using the pattern for}%
\typeout{** the default language instead.}%
\else
\language=\csname l@#1\endcsname
\fi
#2}}
\providecommand{\BIBdecl}{\relax}
\BIBdecl

\bibitem{heemels2012introduction}
W.~{Heemels}, K.~{Johansson}, and P.~{Tabuada}, ``An introduction to
  event-triggered and self-triggered control,'' in \emph{2012 IEEE 51st IEEE
  Conference on Decision and Control}, Dec 2012, pp. 3270--3285.

\bibitem{antunes2014rollout}
D.~J. {Antunes} and W.~P. M.~H. {Heemels}, ``Rollout event-triggered control:
  Beyond periodic control performance,'' \emph{IEEE Transactions on Automatic
  Control}, vol.~59, no.~12, pp. 3296--3311, 2014.

\bibitem{soleymani2017event}
T.~{Soleymani}, S.~{Hirche}, and J.~S. {Baras}, ``Event-triggered
  output-feedback \mbox{$H_\infty$} control with minimum directed
  information,'' in \emph{2017 IEEE 56th Conference on Decision and Control},
  2017, pp. 6088--6094.

\bibitem{nowzari2017event}
C.~Nowzari, E.~Garcia, and J.~Cortes, ``Event-triggered communication and
  control of networked systems for multi-agent consensus,'' \emph{Automatica},
  vol. 105, pp. 1--27, 2019.

\bibitem{Asadi2018Aconsistent}
B.~{Asadi Khashooei}, D.~{Antunes}, and W.~{Heemels}, ``A consistent
  threshold-based policy for event-triggered control,'' \emph{IEEE Control
  Systems Letters}, vol.~2, no.~3, pp. 447--452, July 2018.

\bibitem{wang2011event}
X.~{Wang} and M.~D. {Lemmon}, ``Event-triggering in distributed networked
  control systems,'' \emph{IEEE Transactions on Automatic Control}, vol.~56,
  no.~3, pp. 586--601, 2011.

\bibitem{molin2014bi}
A.~Molin and S.~Hirche, ``A bi-level approach for the design of event-triggered
  control systems over a shared network,'' \emph{Discrete Event Dynamic
  Systems}, vol.~24, no.~2, pp. 153--171, 2014.

\bibitem{mamduhi2014event}
M.~H. {Mamduhi}, D.~{Toli{\'c}}, A.~{Molin}, and S.~{Hirche}, ``Event-triggered
  scheduling for stochastic multi-loop networked control systems with packet
  dropouts,'' in \emph{2014 IEEE 53rd Conference on Decision and Control},
  2014, pp. 2776--2782.

\bibitem{kurose2005computer}
J.~F. Kurose, \emph{Computer networking: A top-down approach featuring the
  internet, 3/E}.\hskip 1em plus 0.5em minus 0.4em\relax Pearson Education
  India, 2005.

\bibitem{Brunner2018stochastic}
F.~D. Brunner, D.~Antunes, and F.~Allg{\"o}wer, ``Stochastic thresholds in
  event-triggered control: A consistent policy for quadratic control,''
  \emph{Automatica}, vol.~89, pp. 376--381, 2018.

\bibitem{han2017stochastic}
D.~{Han}, J.~{Wu}, Y.~{Mo}, and L.~{Xie}, ``On stochastic sensor network
  scheduling for multiple processes,'' \emph{IEEE Transactions on Automatic
  Control}, vol.~62, no.~12, pp. 6633--6640, 2017.

\bibitem{gatsis2016control}
K.~{Gatsis}, A.~{Ribeiro}, and G.~J. {Pappas}, ``Control-aware random access
  communication,'' in \emph{2016 ACM/IEEE 7th International Conference on
  Cyber-Physical Systems (ICCPS)}, 2016, pp. 1--9.

\bibitem{cervin2008scheduling}
A.~{Cervin} and T.~{Henningsson}, ``Scheduling of event-triggered controllers
  on a shared network,'' in \emph{2008 IEEE 47th Conference on Decision and
  Control}, 2008, pp. 3601--3606.

\bibitem{blind2013time}
R.~Blind and F.~Allg{\"o}wer, ``On time-triggered and event-based control of
  integrator systems over a shared communication system,'' \emph{Mathematics of
  Control, Signals, and Systems}, vol.~25, no.~4, pp. 517--557, 2013.

\bibitem{xia2017networked}
M.~Xia, V.~Gupta, and P.~J. Antsaklis, ``Networked state estimation over a
  shared communication medium,'' \emph{IEEE Transactions on Automatic Control},
  vol.~62, no.~4, pp. 1729--1741, 2017.

\bibitem{Ramesh2011ACC}
C.~{Ramesh}, H.~{Sandberg}, L.~{Bao}, and K.~H. {Johansson}, ``On the dual
  effect in state-based scheduling of networked control systems,'' in
  \emph{2011 American Control Conference}, 2011, pp. 2216--2221.

\bibitem{mamduhi2016decentralized}
M.~H. {Mamduhi}, M.~{Kneissl}, and S.~{Hirche}, ``Decentralized event-triggered
  medium access control for networked control systems,'' in \emph{2016 IEEE
  55th Conference on Decision and Control}, 2016, pp. 513--519.

\bibitem{mamduhi2017error}
M.~H. Mamduhi, A.~Molin, D.~Toli{\'c}, and S.~Hirche, ``Error-dependent data
  scheduling in resource-aware multi-loop networked control systems,''
  \emph{Automatica}, vol.~81, pp. 209--216, 2017.

\bibitem{molin:13}
A.~Molin and S.~Hirche, ``On the optimality of certainty equivalence for
  event-triggered control systems,'' \emph{IEEE Transactions on Automatic
  Control}, vol.~58, no.~2, pp. 470--474, Feb 2013.

\bibitem{goldenshluger2017minimum}
A.~{Goldenshluger} and L.~{Mirkin}, ``On minimum-variance event-triggered
  control,'' \emph{IEEE Control Systems Letters}, vol.~1, no.~1, pp. 32--37,
  2017.

\bibitem{lee2006jointly}
{Jang-Won Lee}, M.~{Chiang}, and R.~A. {Calderbank}, ``Jointly optimal
  congestion and contention control based on network utility maximization,''
  \emph{IEEE Communications Letters}, vol.~10, no.~3, pp. 216--218, 2006.

\bibitem{balaghi2018decentralized}
M.~H. {Balaghi I.}, D.~J. {Antunes}, M.~H. {Mamduhi}, and S.~{Hirche}, ``A
  decentralized consistent policy for event-triggered control over a shared
  contention - based network,'' in \emph{2018 IEEE 57th Conference on Decision
  and Control}, Dec 2018, pp. 1719--1724.

\bibitem{antunes2016con}
D.~J. {Antunes} and B.~{Asadi Khashooei}, ``Consistent event-triggered methods
  for linear quadratic control,'' in \emph{2016 IEEE 55th Conference on
  Decision and Control}, 2016, pp. 1358--1363.

\bibitem{Moin2014Thesis}
A.~Molin, ``Optimal event-triggered control with communication constraints,''
  Dissertation, Technical University of Munich, Munich, 2014.

\bibitem{balaghiinaloo2019decentralized}
M.~Balaghiinaloo, D.~J. Antunes, M.~H. Mamduhi, and S.~Hirche, ``Decentralized
  {LQ}-consistent event-triggered control over a shared contention-based
  network,'' \emph{arXiv preprint arXiv:1910.04582}, 2019.

\bibitem{Imer2006optimal}
O.~C. Imer, S.~Y{\"u}ksel, and T.~Ba{\c{s}}ar, ``Optimal control of {LTI}
  systems over unreliable communication links,'' \emph{Automatica}, vol.~42,
  no.~9, pp. 1429--1439, 2006.

\bibitem{schenato2007foundations}
L.~{Schenato}, B.~{Sinopoli}, M.~{Franceschetti}, K.~{Poolla}, and S.~S.
  {Sastry}, ``Foundations of control and estimation over lossy networks,''
  \emph{Proceedings of the IEEE}, vol.~95, no.~1, pp. 163--187, 2007.

\bibitem{balaghi2018optimal}
M.~H. {Balaghi I.}, D.~J. Antunes, M.~H. Mamduhi, and S.~Hirche, ``An optimal
  {LQG} controller for stochastic event-triggered scheduling over a lossy
  communication network,'' in \emph{IFAC-PapersOnLine}, vol.~51, no.~23, 2018,
  pp. 58--63.

\end{thebibliography}
\newpage
\newpage
\newpage
\appendix
\setcounter{equation}{0}
\renewcommand{\theequation}{\thesection.\arabic{equation}}
\subsection{Theorem~\ref{th:decentralizedoptimal}}
  \par Let us consider the whole NCS centrally and denote \mbox{$\mathsf{H}_k=\cup_{i=1}^m \mathsf{H}^i_k$} as the total information available for the hypothetical central controller at every time step. First of all, we have to establish the certainty equivalence property, i.e., that the control law takes the form
\begin{equation}\label{eq:ucentral}
u^i_k=K_i\mathsf{E}[x^i_k|\mathsf{H}_k],
\end{equation}
where due to physical independency between different dynamic users, the optimal control gain for every system can be determined independently, as given in~\eqref{eq:controlgain}. As explained in [Shalom]
, we can conclude the certainty equivalence property if we show the following two features:
\begin{itemize}
  \item Independency of the scheduling law from the control inputs, which prevents the generation of the control inputs' dual effect in the control loops.
  \item Independency of the remote state estimation error from the control inputs.
\end{itemize}
Since every $\delta^i_k$ follows a constant function of the primitive random variables, any \mbox{$\sigma^i_k=\delta^i_k \prod_{j=1,j\neq i}^{m}(1-\delta^j_k)$} is independent from all control inputs, which indicates that the first required property holds. Moreover, following the same steps as in~\cite{Moin2014Thesis}, we can prove that the state estimation errors, i.e. \mbox{$e^i_{k|k}=x_k^i-\mathsf{E}[x^i_k|\mathsf{H}_k]$}, for both forced and unforced dynamics with the same realization of the primitive random variables are equal, therefore, the state estimation error in the controller is independent from the control inputs. Accordingly, the optimal control law for every system follows~\eqref{eq:ucentral}.
\par Now we have to show that \mbox{$\mathsf{E}[x^i_k|\mathsf{H}_k]=\mathsf{E}[x^i_k|\mathsf{H}^i_k]$} holds for all dynamic users at every time step. Based on the Bayes law of conditional probability, we have the following equality at all time steps and for all dynamic users
$$\mathsf{f}(x^i_k|\mathsf{H}^i_k,\cup_{j=1,j\neq i}^{m}\mathsf{H}^j_k)=\frac{\mathsf{Pr}(\cup_{j=1,j\neq i}^{m}\mathsf{H}^j_k|\mathsf{H}^i_k,x^i_k)}{\mathsf{Pr}(\cup_{j=1,j\neq i}^{m}\mathsf{H}^j_k|\mathsf{H}^i_k)} \mathsf{f}(x^i_k|\mathsf{H}^i_k),$$
where it is clear that the fraction term in the above equation is equal to one and
\mbox{$\mathsf{f}(x^i_k|\mathsf{H}^i_k,\cup_{j=1,j\neq i}^{m}\mathsf{H}^j_k)= \mathsf{f}(x^i_k|\mathsf{H}^i_k)$}, which concludes the statement. Therefore, the optimal local controllers~\eqref{eq:u_optimal} are equivalent to the optimal central controller. 
\subsection{Corollary \ref{lem:PSTestimator}}
We assume that the local estimator is aware of the control law and since \mbox{$\mathsf{H}^i_{k-1}\subseteq \mathsf{L}^i_k$}, then the local estimator is aware of all the previous control input values, i.e., \mbox{$\{u^i_{\ell}|\ell<k\}$} at every time step $k$. As discussed before, the information set of the local and the remote estimator is equivalent at triggering time steps, therefore, \mbox{$\bar{x}^i_{k|k}=\hat{x}^i_{k|k}$}, when \mbox{$\sigma^i_k=1$}. However, when $\sigma^i_k=0$, then the remote state estimation pdf is as follows
$$\begin{aligned}
\mathsf{f}(x^i_k|\mathsf{H}^i_{k-1},\sigma^i_k=0)=\frac{\mathsf{Pr}(\sigma^i_k=0|\mathsf{H}^i_{k-1},x^i_k)}{\mathsf{Pr}(\sigma^i_k=0|\mathsf{H}^i_{k-1})}\mathsf{f}(x^i_k|\mathsf{H}^i_{k-1}).
\end{aligned}$$
Since the unsuccessful transmission probability is independent from $x^i_k$, the fraction term in the right hand side of the above equation is equal to one and
\mbox{$\mathsf{f}(x^i_k|\mathsf{H}^i_{k-1},\sigma^i_k=0)=\mathsf{f}(x^i_k|\mathsf{H}^i_{k-1})$},
which indicates that in case of unsuccessful transmission, it is just needed to perform the prediction stage of the Kalman filter, i.e., \mbox{$\bar{x}^i_{k|k}=\bar{x}^i_{k|k-1}$}, where \mbox{$\bar{x}^i_{k+1|k}=A_i\bar{x}^i_{k|k}+B_iu^i_k$}.
Therefore, the conditional state expectation in the remote estimator is proved to be given by~\eqref{eq:Observer}.
\subsection{Lemma~\ref{lem:PSTcost}}
\par We can represent the Kalman filter as
\begin{equation}\label{eq:scheduler-estimate-p}
 \hat{x}^i_{k+1|k+1}=A_i\hat{x}^i_{k|k}+B_iu^i_k+L_i(y^i_{k+1}-C_i\hat{x}^i_{k+1|k}),
\end{equation}
and the remote state estimator as
\begin{equation}\label{eq:controller-estimate-p}
 \bar{x}^i_{k+1|k}=\sigma^i_k A_i\hat{x}^i_{k|k}+(1-\sigma^i_k) A_i\bar{x}^i_{k|k-1}+B_iu^i_k.
 \end{equation}
By subtracting \eqref{eq:controller-estimate-p} from \eqref{eq:scheduler-estimate-p}, the dynamics of the predicted error (\mbox{$e^i_{k|k-1}=\hat{x}^i_{k|k}-\bar{x}^i_{k|k-1}$}) is as given in~\eqref{eq:schedulerserror} for \mbox{$\hat{e}^i_{k|k-1}=x^i_{k}-\hat{x}^i_{k|k-1}$} and \mbox{$\hat{e}^i_{k|k}=x^i_{k}-\hat{x}^i_{k|k}$}, where 
\begin{equation}\label{eq:estimate-error2-p}
 \begin{aligned}
 &\hat{e}^i_{k+1|k}=A_i(I-L_iC_i)\hat{e}^i_{k|k-1}-A_iL_iv^i_k+w^i_k,&\\&
 \hat{e}^i_{k+1|k+1}=A_i\hat{e}^i_{k|k}+w^i_k-L_i(C_i\hat{e}^i_{k+1|k}+v^i_{k+1}).
 \end{aligned}
 \end{equation}
We know that when $\sigma^i_k=1$, then the updated remote state estimation error is \mbox{$\bar{e}^i_{k|k}=x^i_k-\bar{x}^i_{k|k}=\hat{e}^i_{k|k}$}. However, when \mbox{$\sigma^i_k=0$}, then
$\bar{e}^i_{k|k}=x^i_k-\hat{x}^i_{k|k}+\hat{x}^i_{k|k}-\bar{x}^i_{k|k-1}=\hat{e}^i_{k|k}+e^i_{k|k-1}$.
Using \eqref{eq:schedulerserror} and \eqref{eq:estimate-error2-p}, \mbox{$\bar{e}_{k|k}$} will have the following dynamics
\begin{equation}\label{eq:errordynamic1}
  \bar{e}^i_{k+1|k+1}=A_i\bar{e}^i_{k|k}+w^i_{k}
\end{equation}
Therefore, we can express the updated remote state estimation error dynamics as
\begin{equation}\label{eq:errordynamic2}
  \bar{e}^i_{k+1|k+1}=\begin{cases}
                      A_i\hat{e}^i_{k|k}+w^i_{k}, & \mbox{if } \sigma^i_{k}=1 \\
                      A_i\bar{e}^i_{k|k}+w^i_k, & \mbox{otherwise}.
                    \end{cases}
\end{equation}
Denoting by \mbox{$\Phi^i_{k|k}=\mathsf{E}[\bar{e}^i_{k|k}\bar{e}^{i\intercal}_{k|k}|\mathsf{H}^i_k]$} the covariance of the updated remote state estimation error at every time step $k$, based on~\eqref{eq:errordynamic2}, we can conclude that it has the following dynamics
$$\Phi^i_{k+1|k+1}=(1-\sigma^i_{k+1})(A_i \Phi^i_{k|k}A_i^\intercal+W^i)+\sigma^i_{k+1} \Theta_i.$$
From the fact that \mbox{$\sigma^i_{k+1}$} is independent from \mbox{$\Phi^i_{k|k}$}, we obtain
$$\mathsf{E}[\Phi^i_{k+1|k+1}]=(1-q^ip^i)(A_i \mathsf{E}[\Phi^i_{k|k}]A^\intercal_i+W_i)+q^ip^i \Theta_i.$$
Moreover, letting \mbox{$\bar{\Phi}^i=\limsup_{k\rightarrow \infty}\mathsf{E}[\Phi^i_{k|k}]$}, we find
$$\bar{\Phi}^i=(1-q^ip^i)(A_i \bar{\Phi}^iA^\intercal_i+W_i)+q^ip^i \Theta_i.$$
This equation has the following closed form solution
$$\bar{\Phi}^i=\sum_{j=0}^{\infty} (1-q^ip^i)^{j+1}(A^j_i W_iA^{j\intercal}_i)+q^ip^i(1-q^ip^i)^{j}(A^{j}_i \Theta_i A^{j\intercal}_i),$$
which is bounded if \mbox{$\varrho(\sqrt{1-q^ip^i}A_i)<1$}, see \cite{Imer2006optimal,schenato2007foundations}. On the other hand, the average quadratic performance~\eqref{eq:per} is given by [Bertsekas] 
as follows
\begin{equation}\label{eq:perfIndex2}
J^i=\text{tr}(P_iW_i) +\limsup_{T\rightarrow \infty}\frac{1}{T}\sum_{k=0}^{T-1}\text{tr}(Y_i\mathsf{E}[\Phi^i_{k|k}]),
\end{equation}
which can be expressed as
$$\begin{aligned}
J^i&=\text{tr}(P_iW_i) +\text{tr}(Y_i\limsup_{k\rightarrow \infty}\mathsf{E}[\Phi^i_{k|k}])=\text{tr}(P_iW_i)+\text{tr}(Y_i\bar{\Phi}^i).
\end{aligned}$$
By substituting the solution of $\bar{\Phi}^i$ in the above equation, we arrive at the closed form of the average control performance for the PST policy in \eqref{eq:pscost}.
\subsection{Lemma \ref{le:twogaussians}}
\par We consider both situations separately and drop the index~$i$ for simplicity.
\par 1) \mbox{$\delta^{st}_k=0$}:  based on the Bayes law of conditional probability we have
\begin{equation}\label{eq:N1}
      \begin{aligned}
    \mathsf{f}(e_{k|k}&|\delta^{st}_k=0, \mathsf{T}_{k})=\frac{\mathsf{Pr}(\delta^{st}_{k}=0|e_{k|k-1},\mathsf{T}_{k})}{\mathsf{Pr}(\delta^{st}_{k}=0|\mathsf{T}_{k})}\mathsf{f}(e_{k|k-1}| \mathsf{T}_{k}).
\end{aligned}
    \end{equation}
Denote \mbox{$z:=e_{k|k-1}$}, then based on the triggering policy~\eqref{eq:scheduler2} and considering \mbox{$r_0:=\frac{1}{2}z^{\intercal}\Psi^{-1}_{k|k-1}z$}, we have
\begin{equation}\label{eq:Psige}
    \begin{aligned}
\mathsf{Pr}(\delta^{st}_{k}=0|z,\mathsf{T}_{k})\!=\!\int\limits_{r_0}^\infty \lambda_ke^{-\lambda_k r}dr=e^{-\frac{\lambda_k}{2}z^{\intercal}\Psi^{-1}_{k|k-1}z},\end{aligned}
  \end{equation}
  and
   \begin{equation}\label{eq:psig}
      \begin{aligned}
\mathsf{Pr}(\delta^{st}_{k}=0|\mathsf{T}_{k})&=\int\limits_{z\in\mathbb{R}^{n}}\int\limits_{r_{0},}^{\infty}\frac{(\lambda_ke^{-\lambda_kr})e^{-\frac{1}{2}z^\intercal\Psi^{-1}_{k|k-1}  z}}{\text{det}(2\pi\Psi_{k|k-1})^\frac{1}{2}}drdz&\\&=\int\limits_{z\in\mathbb{R}^{n}}\frac{e^{-\frac{1+\lambda_k}{2}z^\intercal\Psi^{-1}_{k|k-1}z}}{\text{det}(2\pi\Psi_{k|k-1})^\frac{1}{2}}dz=(1+\lambda_k)^{-\frac{n}{2}}.\end{aligned}
    \end{equation}
Finally, by substituting~\eqref{eq:Gaussian},~\eqref{eq:Psige} and~\eqref{eq:psig} into~\eqref{eq:N1},
$$\begin{aligned}
    \mathsf{f}(e_{k|k}|\delta^{st}_{k}=0, \mathsf{T}_{k})&=\frac{e^{-\frac{1+\lambda_k}{2}z^\intercal\Psi^{-1}_{k|k-1}  z}}{(1+\lambda_k)^{-\frac{n}{2}}\text{det}(2\pi\Psi_{k|k-1})^\frac{1}{2}}&\\&=\mathsf{N}(0,\frac{\Psi_{k|k-1}}{1+\lambda_k}).
\end{aligned}$$
Therefore, when \mbox{$\delta^{st}_k=0$}, the pdf of the updated state estimation error \mbox{$e_{k|k}$} will remain Gaussian. Moreover, since \mbox{$\nu_{k+1}= L(C\hat{e}_{k+1|k}+v_{k+1})$} in the dynamics of the predicted state estimation error~\eqref{eq:schedulerserror} is Gaussian, then the predicted state estimation error at the next time step is also Gaussian as given in~\eqref{eq:nottriggeringpdf}, where
\mbox{$\Phi= A\Theta A^\intercal -\Theta+W=\mathsf{E}[\nu_{k+1}\nu_{k+1}^\intercal]$} for all $k$.
\par 2) \mbox{$\delta^{st}_k=1,~\rho_k=0$}: in this case, the controller does not receive $x_k$, therefore,
\begin{equation}\label{eq:C1}
  \begin{aligned}
\mathsf{f}(&e_{k|k}|\delta^{st}_k=1,\rho_k=0, \mathsf{T}_{k})=&\\&\frac{\mathsf{Pr}(\delta^{st}_k=1|e_{k|k-1},\rho_k=0,\mathsf{T}_{k})}{\mathsf{Pr}(\delta^{st}_k=1|\rho_k=0,\mathsf{T}_{k})}\mathsf{f}(e_{k|k-1}| \mathsf{T}_{k}).
\end{aligned}
\end{equation}
By using \eqref{eq:Psige} and \eqref{eq:psig}, we get
\begin{equation}\label{eq:sigma1}
    \begin{aligned}
\mathsf{Pr}&(\delta^{st}_k=1|z,\rho_k=0,\mathsf{T}_{k})=1-e^{-\frac{\lambda_k}{2}z^{\intercal}\Psi^{-1}_{k|k-1}z},
\end{aligned}
  \end{equation}
  and
\mbox{$p_k:=\mathsf{Pr}(\delta^{st}_k=1|\rho_k=0,\mathsf{T}_{k})=1-(1+\lambda_k)^{-\frac{n}{2}}$}. Then by substitution into \eqref{eq:C1}, we get
$$\begin{aligned}
\mathsf{f}(e_{k|k}|\delta^{st}_k=1,\rho_k=0, \mathsf{T}_{k})=\frac{1}{p_k}&\big(\frac{e^{-\frac{1}{2}z^\intercal\Psi^{-1}_{k|k-1}z}}{\text{det}(2\pi\Psi_{k|k-1})^\frac{1}{2}}
&\\&-\frac{e^{-\frac{1+\lambda_k}{2}z^\intercal\Psi^{-1}_{k|k-1}z}}{\text{det}(2\pi\Psi_{k|k-1})^\frac{1}{2}}\big).
\end{aligned}$$
Therefore, in case of a data collision, the pdf of the updated state estimation error will become the sum of two Gaussians
$$\begin{aligned}
\mathsf{f}(e_{k|k}|\delta^{st}_k=1,\rho_k=0, \mathsf{T}_{k})&=\frac{1}{p_k}\mathsf{N}(0,\Psi_{k|k-1})&\\&-\frac{1-p_k}{p_k}\mathsf{N}(0,\frac{\Psi_{k|k-1}}{1+\lambda_k}).
\end{aligned}$$
Again with the same conclusion as the one presented for the case when \mbox{$\delta^{st}_k=0$}, the predicted state estimation error at the next time step will be the sum of two Gaussians, where their covariances follow~\eqref{eq:updatedcovariance}.

\subsection{Theorem \ref{th:opt}}
\par First of all, we have to show that the CETT or equivalently STETT policy is in the class of admissible schedulers. The Kalman filter by the scheduler side and the remote state estimator follow~\eqref{eq:scheduler-estimate-p} and~\eqref{eq:controller-estimate-p}, respectively.
By subtracting~\eqref{eq:controller-estimate-p} from~\eqref{eq:scheduler-estimate-p}, the dynamics of the state estimation error used by the scheduling policy~\eqref{eq:scheduler2} is determined by~\eqref{eq:schedulerserror}.
Then, the scheduling law~\eqref{eq:scheduler2} can be represented as
\mbox{$\delta^i_k=g(\mathsf{R}^i_k)$},
where $g:\mathbb{R}^n\times \mathbb{R}^{nk}\times\mathbb{R}^{o(k+1)}\times\mathbb{R}^{r(k+1)}\rightarrow \{0,1\}$ is an appropriate function and
\mbox{$\mathsf{R}^i_k=\{v^i_t,w^i_t,r^i_t,\rho^i_t|t\in Z^{k-1}_0\}\cup\{v^i_{k}\} \cup\{x^i_0\}$} is a set of independent primitive random variables. Therefore, the CETT policy is in the class of admissible schedulers and the certainty equivalent controller is optimal based on Theorem~\ref{th:decentralizedoptimal}.
\par Now we have to find the estimated state in the controller for which we follow an induction arrangement. For the sake of simplicity, we consider a single control loop and drop the index $i$. Without loss of generality, let us assume that at \mbox{$k=0$}, \mbox{$\sigma_0=1$} and find the state estimation at \mbox{$k=1$}, assuming \mbox{$\sigma_1=0$}. Then,
\mbox{$\mathsf{f}(x_0|\sigma_0=1,\hat{x}_{0|0})=\mathsf{N}(\hat{x}_{0|0},\Theta)$},
which can be concluded based on the properties of the Kalman filter. Since the remote state estimator is aware of the control inputs at all time steps, the pdf of the predicted state at \mbox{$k=1$} is
\begin{equation}\label{eq:predictionstage}
  \mathsf{f}(x_1|\sigma_0=1,\hat{x}_{0|0})=\mathsf{N}(\bar{x}_{1|0},\Gamma_{1|0}),
\end{equation}
where \mbox{$\bar{x}_{1|0}=A\hat{x}_{0|0}+Bu_0$} and \mbox{$\Gamma_{1|0}=A\Theta A^\intercal +W$}. Then the updated pdf of the remote state estimation at \mbox{$k=1$} if \mbox{$\sigma_1=0$} is determined by using Bayes law of conditional probability as follows
\begin{equation}\label{eq:unsuccessfultransmission1}
\begin{aligned}
\mathsf{f}(x_1|&\sigma_0=1,\sigma_1=0,\hat{x}_{0|0})&\\&=\frac{\mathsf{Pr}(\sigma_1=0|\sigma_0=1,\hat{x}_{0|0},x_1)}{\mathsf{Pr}(\sigma_1=0|\sigma_0=1,\hat{x}_{0|0})}\mathsf{f}(x_1|\sigma_0=1,\hat{x}_{0|0}).
\end{aligned}
\end{equation}
Moreover, we have
\begin{equation}\label{eq:prob2}
\begin{aligned}
\mathsf{Pr}(\sigma_1=0|&\sigma_0=1,\hat{x}_{0|0},x_1)=\mathsf{Pr}(\delta_1=0|\sigma_0=1,\hat{x}_{0|0},x_1)&\\&+\mathsf{Pr}(\rho_1=0)\mathsf{Pr}(\delta_1=1|\sigma_0=1,\hat{x}_{0|0},x_1).
\end{aligned}
\end{equation}
Let \mbox{$\hat{z}_1:=x_1-\hat{x}_{1|1},$} \mbox{$\bar{z}_1:=x_1-\bar{x}_{1|0}$} and \mbox{$z_1:=\hat{x}_{1|1}-\bar{x}_{1|0}$},
where
\mbox{$\mathsf{f}(\hat{z}_1)=\mathsf{N}(0,\Theta)$} and \mbox{$\mathsf{E}[z_1z_1^\intercal|\sigma_0=1]=\Psi_{1|0}$}
for \mbox{$\Psi_{1|0}=\Phi=A\Theta A^\intercal-\Theta+W$}. Then
$$\begin{aligned}
&\mathsf{Pr}(\delta_1=0|\sigma_0=1,\hat{x}_{0|0},x_1)=\mathsf{Pr}(\delta_1=0|\sigma_0=1,\bar{x}_{1|0},x_1)=&\\&\mathsf{Pr}(\delta_1=0|\sigma_0=1,\bar{z}_1)
=\int\limits_{\hat{z}_1\in\mathbb{R}^{n}}\int\limits_{r_{0},}^{\infty}\frac{\lambda_1e^{-\lambda_1r}e^{-\frac{1}{2}\hat{z}_1^\intercal\Theta^{-1}  \hat{z}_1}}{\text{det}(2\pi\Theta)^\frac{1}{2}}drd\hat{z}_1
&\\&=\int\limits_{\hat{z}_1\in\mathbb{R}^{n}}\frac{e^{-\frac{1}{2}(\bar{z}_1-\hat{z}_1)^\intercal \lambda\Psi^{-1}_{1|0}  (\bar{z}_1-\hat{z}_1)-\frac{1}{2}\hat{z}_1^\intercal\Theta^{-1}  \hat{z}_1}}{\text{det}(2\pi\Theta)^\frac{1}{2}}d\hat{z}_1,
\end{aligned}$$
for \mbox{$r_0:=\frac{1}{2}z^{\intercal}_1\Psi^{-1}_{1|0}z_1=\frac{1}{2}(\bar{z}_1-\hat{z}_1)^{\intercal}\Psi^{-1}_{1|0}(\bar{z}_1-\hat{z}_1)$}. We can simplify the above equation by using the following equality
$$\begin{aligned}
(\bar{z}_1&-\hat{z}_1)^\intercal\lambda_1\Psi^{-1}_{1|0}(\bar{z}_1-\hat{z}_1)+\hat{z}_1^\intercal\Theta^{-1}\hat{z}_1&\\&
=(\hat{z}_1-\bar{z}_1^\prime)^\intercal(\lambda_1\Psi^{-1}_{1|0}+\Theta^{-1})(\hat{z}_1-\bar{z}_1^\prime)+\bar{z}_1^\intercal\Pi_{1|0}^{-1}\bar{z}_1,
\end{aligned}$$
where
$\bar{z}^\prime_1= (\lambda_1\Psi^{-1}_{1|0}+\Theta^{-1})^{-1}\lambda_1\Psi^{-1}_{1|0}\bar{z}_1,$ and
$$
\begin{aligned}
\Pi_{1|0}&=\big(\lambda_1 \Psi^{-1}_{1|0}-\lambda_1 \Psi^{-1}_{1|0}(\lambda_1 \Psi^{-1}_{1|0}+\Theta^{-1})^{-1}\lambda_1 \Psi^{-1}_{1|0}\big)^{-1}
&\\&=\frac{1}{\lambda_1} \Psi_{1|0}+\Theta.
\end{aligned}
$$
Then
\begin{equation}\label{eq:pro3}
  \mathsf{Pr}(\delta_1=0|\sigma_0=1,\hat{x}_{0|0},x_1)=\xi_1 e^{-\frac{1}{2}\bar{z}^\intercal_1 \Pi^{-1}_{1|0} \bar{z}_1},
\end{equation}
where
$\xi_1=1/\big(\det (\lambda_1 \Psi^{-1}_{1|0}+\Theta^{-1})\det(\Theta)\big).$
Moreover,
\begin{equation}\label{eq:pro4}
  \mathsf{Pr}(\delta_1=1|\sigma_0=1,\hat{x}_{0|0},x_1)=1-\xi_1 e^{-\frac{1}{2}\bar{z}_1^\intercal \Pi^{-1}_{1|0} \bar{z}_1}.
\end{equation}
Now substitute \eqref{eq:pro3} and \eqref{eq:pro4} into \eqref{eq:prob2}, which results in
$$\begin{aligned}
\mathsf{Pr}(\sigma_1=0|\sigma_0=1,\hat{x}_{0|0},x_1)=1-q_1+q_1\xi_1 e^{-\frac{1}{2}\bar{z}^\intercal_1 \Pi^{-1}_{1|0} \bar{z}_1},
\end{aligned}$$
where \mbox{$q_1:=\mathsf{Pr}(\rho_1=1)$}, then substitute the result into~\eqref{eq:unsuccessfultransmission1}, which results in the following
\begin{equation}\label{eq:unsuccessfultransmission2}
\begin{aligned}
\mathsf{f}(x_1|&\sigma_0=1,\sigma_1=0,\hat{x}_{0|0})&\\&=\frac{1-q_1+q_1\xi_1 e^{-\frac{1}{2}\bar{z}_1^\intercal \Pi^{-1}_{1|0} \bar{z}_1}}{1-q_1p_1}
\frac{e^{-\frac{1}{2}\bar{z}_1^\intercal \Gamma^{-1}_{1|0}\bar{z}_1}}{\det(2\pi \Gamma_{1|0})^\frac{1}{2}}&\\&
=\sum_{i=1}^{2}\beta^\prime_i \mathsf{N}(\bar{x}_{1|0},\Gamma^{i}_{1|1}),
\end{aligned}
\end{equation}
where $\sum_{i=1}^{2}\beta^\prime_i=1$. As it can be seen, at the first time step after a successful transmission the updated state estimation pdf is the sum of two Gaussian terms with different covariances. The total covariance of the estimation is affected by $q_1$, which is the probability that the network is available. However, the mean values of the Gaussian terms are equal to the one obtained at the prediction stage~\eqref{eq:predictionstage} and do not depend on $q_1$, i.e.,
\mbox{$\bar{x}_{1|1}=\mathsf{E}[x_1|\sigma_0=1,\sigma_1=0,\hat{x}_{0|0}]=\bar{x}_{1|0}$}.
Now assume \mbox{$\sigma_k=0,~\forall k\in Z_1^{t+1}$}, then the updated state estimation at \mbox{$k=t+1$} is determined as follows
\begin{equation}\label{eq:pdfprobability}
\begin{aligned}
\mathsf{f}(x_{t+1}&|\nu_t,\sigma_{t+1}=0,\hat{x}_{0|0})&\\&=\frac{\mathsf{Pr}(\sigma_{t+1}=0|\nu_t,\hat{x}_{0|0},x_{t+1})}{\mathsf{Pr}(\sigma_{t+1}=0|\nu_t,\hat{x}_{0|0})}\mathsf{f}(x_{t+1}|\nu_t,\hat{x}_{0|0}),
\end{aligned}
\end{equation}
where
\mbox{$\nu_t=\{\sigma_0=1,\sigma_1=\dots=\sigma_t=0\}$}.
Let us define
\mbox{$c=\inf \{k|\delta_k=1\land k\in Z_1^t\}$}
as the first time step after the last successful transmission, where data collision happens, otherwise, \mbox{$c=0$}. Then we can partition the set \mbox{$\nu_t$} into several mutually exclusive sets as \mbox{$\nu^c_t$}, where for every \mbox{$c\in Z_1^t$},
$$
\begin{aligned}
\nu_t^c=\big\{\sigma_0=1, \delta_1=\dots=\delta_{c-1}=0, &\{\delta_c=1, \rho_c=0\},&\\& \sigma_{c+1}=\dots=\sigma_t=0\big\},
\end{aligned}$$
and $\nu_t^0=\{\sigma_0=1,\delta_1=\dots=\delta_{t}=0\}.$
Therefore,
\begin{equation}\label{eq:probabilitydevision}
  \begin{aligned}
&\mathsf{Pr}(\sigma_{t+1}=0|\nu_t,\hat{x}_{0|0},x_{t+1})f(x_{t+1}|\nu_t,\hat{x}_{0|0})&\\&=
\sum_{j=0}^{t}\mathsf{Pr}(\sigma_{t+1}=0|\nu^j_t,\hat{x}_{0|0},x_{t+1})\mathsf{Pr}(c=j)\mathsf{f}(x_{t+1}|\nu^j_t,\hat{x}_{0|0}).
\end{aligned}
\end{equation}
 According to the operation mechanism of the CETT policy, the predicted state estimation has the following distribution
$$\mathsf{f}(x_{t+1}|\nu^j_t,\hat{x}_{0|0})=\begin{cases}
                                     \mathsf{N}(\bar{x}_{t+1|t},\Gamma^{0}_{t+1|t}), & \mbox{if } j= 0 \\
                                     \sum_{i=1}^{2}\mathsf{N}(\bar{x}_{t+1|t},\Gamma^{j,i}_{t+1|t}), & \mbox{otherwise,}
                                   \end{cases}$$
where according to the induction assumption, the mean values of all Gaussian terms are equal and determined as \mbox{$\bar{x}_{t+1|t}=A\bar{x}_{t|t-1}+Bu_t$} and \mbox{$\Gamma_{t+1|t}^{j,i}=A\Gamma_{t|t}^{j,i} A^\intercal +W$} is their covariance.
\newline Moreover, based on the triggering policy, after the first collision instance, which results in two Gaussian terms in the pdf of the state estimation, the scheduling policy switches to the purely stochastic policy, where
\mbox{$\mathsf{Pr}(\delta_{t+1}=0|\nu^j_t,\hat{x}_{0|0},x_{t+1})=1-p_{t+1},~\forall j\neq 0.$}
However, following the same procedure as the one for \mbox{$t=1$},
$$\begin{aligned}
\mathsf{Pr}(\sigma_{t+1}=0|\nu^0_t,\hat{x}_{0|0},x_{t+1})&=1-q_{t+1}&\\&+q_{t+1}\xi_{t+1} e^{-\frac{1}{2}\bar{z}^\intercal_{t+1} \Pi^{-1}_{t+1|t} \bar{z}_{t+1}},
\end{aligned}$$
where \mbox{$\xi_{t+1}=1/\big(\det (\lambda_{t+1} \Psi^{-1}_{t+1|t}+\Theta^{-1})\det(\Theta)\big)$}, $\bar{z}_{t+1}=x_{t+1}-\bar{x}_{t+1|t}$ and \mbox{$\Pi_{t+1|t}=\frac{1}{\lambda_{t+1}}\Psi_{t+1|t}+\Theta$}.
Then by substituting the last two expressions into~\eqref{eq:probabilitydevision} and then into~\eqref{eq:pdfprobability},
$$\begin{aligned}
\mathsf{f}(x_{t+1}&|\nu_t,\sigma_{t+1}=0,\hat{x}_{0|0})&\\&=\frac{\sum_{j=0}^{t}\sum_{i=1}^{2}d(i,j,p_{t+1},q_{t+1})\mathsf{N}(\bar{x}_{t+1|t},\Gamma^{j,i}_{t+1|t+1})}{1-q_{t+1}p_{t+1}},
\end{aligned}$$
where \mbox{$d(i,j,p_{t+1},q_{t+1})$} is a scalar function. Therefore, at \mbox{$k=t+1$} the number of Gaussian terms is equal to \mbox{$2(t+1)$}. However, the mean of all these terms are equal and not affected by the kind of scheduling policy (PST or STETT), the triggering probability \mbox{$p_{t+1}$}, or the collision probability \mbox{$q_{t+1}$}, and that is in line with the induction assumption. Therefore, \mbox{$\bar{x}_{t+1|t+1}=\bar{x}_{t+1|t}$} when \mbox{$\sigma_{t+1}=0$} and~\eqref{eq:Observer} still holds, when the scheduler is operating based on CETT policy and the result follows.
\subsection{Proof of Proposition~\ref{Prop:11}}
This proposition actually considers the propagation of the first time step's state estimation error covariance in the future time steps during every transmission epoch.
We know that when \mbox{$\sigma_t=0$} for \mbox{$t\in Z_1^{v-1}$} during every transmission epoch, \mbox{$\bar{e}_{t|t}=\hat{e}_{t|t}+e_{t|t}$}, where \mbox{$\hat{e}_{t|t}\sim \mathsf{N}(0,\Theta)$}.
\par For the PST policy, we know that \mbox{$\bar{e}_{1|1}\sim \mathsf{N}(0,\Psi^{ps}_{1|1}+\Theta)$} which will increase the covariance of the future errors as \mbox{$A^{t-1} (\Psi^{ps}_{1|1}+ \Theta) A^{\intercal t-1}$} for all \mbox{$t\in Z_2^{v-1}$}. Therefore, the total amount of increase of the cost function during every transmission epoch due to the first time step state estimation error will be
\mbox{$\delta \hat{J}_{ps}=\sum_{j=2}^{v-1}\text{tr}\big(A^{ j-1}YA^{\intercal j-1} (\Psi^{ps}_{1|1}+ \Theta)\big)$}, which results in
\mbox{$L_{ps}=\sum_{j=1}^{v-2}A^{j}YA^{\intercal  j}$}.
Now let us consider the CETC policy and denote
$$
\beta(l)=\begin{cases}
           (1-p)^{\frac{2}{n}}, & \mbox{if } l<0 \\
           \frac{1-(1-p)^{1+\frac{2}{n}}}{p}, & \mbox{if } l=0 \\
           1, & \mbox{otherwise}.
         \end{cases}
$$
Suppose that at \mbox{$t=1$} the first collision has occurred. Then from the next time step, the scheduler follows the PST policy, where the increase in the value of the covariance will be as the one obtained for the PST policy, i.e., \mbox{$\delta \hat{J}^{c=1}_{\mu}=\sum_{j=2}^{v-1}\text{tr}\big(A^{ j-1}YA^{\intercal j-1} (\Psi^{\mu}_{1|1}+ \Theta)\big)$} which results in
\mbox{$L^{c=1}_{\mu}=\sum_{j=1}^{v-2}A^{ j}YA^{\intercal j}.$}
Now suppose that collision occurs at \mbox{$t=2$}, then we can show that
\mbox{$L^{c=2}_{\mu}=\sum_{j=1}^{v-2}\beta(0) A^{ j}YA^{\intercal  j}$}
and if collision occurs at \mbox{$t=k>2$}, then
$$L^{c=k}_{\mu}=\sum_{j=1}^{v-2}\prod_{l=2-k}^{2-k+j-1}\beta(l) A^{\intercal j}YA^j.$$
Therefore,
$$L_\mu=\sum_{k=1}^{v-1}p(1-p)^{k-1}L_{c=k}^{\mu}=\sum_{j=1}^{v-2}\alpha(j)A^{ j}YA^{\intercal j},$$
where
\mbox{$\alpha(j)=p+\sum_{k=2}^{v-1}p(1-p)^{k-1}\prod_{l=2-k}^{2-k+j-1}\beta(l)$}.
In order to prove \mbox{$L_\mu\leqslant L_{ps}$}, it is just needed to prove \mbox{$\alpha(j)\leqslant 1$} for all \mbox{$j\in Z_1^{v-2}$}. For an arbitrary \mbox{$j$}, we have
$$\begin{aligned}
\alpha (j)&=p+\beta(0)p(1-p)+\beta(-1)\beta(0) p(1-p)^{2}+\dots&\\&
+\beta(-1)^{j-1}\beta(0)p(1-p)^j+\beta(-1)^jp(1-p)^{j+1}&\\&
+\dots+\beta(-1)^jp(1-p)^{v-2}
=\dots&\\&
=1-(1-p)^{v+\frac{2j}{n}-2}<1,
\end{aligned}$$
which concludes our statement and proves the proposition.
\subsection{Solving the optimization problem in~\eqref{eq:etaoptimization}}
We know \mbox{$\eta^i=p^i\prod_{j=1,j\neq i}^{m}(1-p^j)$}, then
$$\begin{aligned}
\sum_{j=1}^{m}U^j(\eta^j)&=\sum_{j=1}^{m}\big(\log (p^j)^{c^j}+\sum_{i=1,i\neq j}^{m}\log (1-p^i)^{c^j}\big)&\\&
=\sum_{j=1}^{m}\log \big((p^j)^{c^j}(1-p^j)^{\sum_{i=1,i\neq j}^{m} c^i} \big),
\end{aligned}$$
which indicates that
$$p^{j*}=\arg \max_{p^j\in[0,1]} (p^j)^{c^j}(1-p^j)^{\sum_{i=1,i\neq j}^{m} c^i},$$
which results in \mbox{$p^{j*}=c^j/\sum_{i=1}^{m}c^i$}.

\end{document}